\documentclass[10pt]{article}
\usepackage{amsmath}
\usepackage{amsfonts}
\usepackage{array}
\usepackage{caption}
\usepackage{comment}
\usepackage{graphicx}% use this package if an eps figure is included.
\usepackage[a4paper, portrait, margin=2cm]{geometry}
\usepackage{hyperref}
\usepackage{indentfirst} % indent the first paragraph after section header
\usepackage{lineno,hyperref}
\usepackage{mathrsfs}
\usepackage{multirow}
\usepackage{ragged2e}
\usepackage{rotating}
\usepackage{siunitx}
\usepackage{soul}
\usepackage{subcaption}
\usepackage{tabularx}
\usepackage[normalem]{ulem}
\usepackage{url}
\usepackage{xcolor}

\newcommand{\bsym}[1]{\boldsymbol{#1}}

\newcommand{\pow}[2]{#1 \cdot 10^{#2}}

\begin{document}
	
	\centering \Huge Modeling inertial flows with meshless Lattice Boltzmann Method
	
	\vspace{0.75cm}
	\centering
	\large Dawid Strzelczyk$^{*}$, Maciej Matyka \normalsize
	
	\vspace{0.25cm}
	\textit{Institute of Theoretical Physics, Faculty of Physics and Astronomy, University of Wrocław, pl. M. Borna 9, 50-204, Wrocław, Poland}
	
	\vspace{0.25cm}
	\RaggedRight
	$^*$ \texttt{dawid.strzelczyk@uwr.edu.pl}
	\linebreak
	
	\justifying
	
	\noindent \textbf{Abstract} 
	One of the limitations of the Lattice Boltzmann Method in simulating inertial flows is the coupling of the discretization of space to the velocity discretization. It requires an increase of the size of computational lattices in order to increase the Reynolds number at a fixed velocity and viscosity. In this work, we adopt the recently proposed meshless formulation of Lattice Boltzmann Method to the problem of inertial flows. In contrast to the standard algorithm, it allows to decouple space and velocity discretizations. Thus, one can change the conversion factors from lattice to physical units for length, velocity, and body force by scaling the streaming distance. In turn, one increases the Reynolds number without increasing the size of the discretization. We measure the accuracy and efficiency of this approach in the K\'{a}rm\'{a}n vortex street behind a circular obstacle and the flow through a porous sample in Darcy and inertial regime. Additionally, we apply the meshless streaming step to the recently proposed fixed relaxation time $\tau=1$ LBM to extend its applicability to model inertial flows.
	\vspace{0.25cm}
	
	\vspace{1cm}
	
	\section{Introduction}
	
%	The transport of fluids through complex, porous structures is ubiquitous with many technological applications (CO2 storage, oil and shale gas extraction, electric battery design). It is of great interest for the study of human health, e.g. blood and cerebrospinal fluid flow, gas transport in the lungs\cite{Ku1997, Kelley2023}.
%	
%	Experimental investigation of fluid flow phenomena in porous media is costly and requires sophisticated setups, nevertheless methods like Particle Image Velocimetry (PIV) has become popular for measuring velocity fields in creeping flows\cite{Morad09, Souzy2020}. However, beyond Darcy Law (inertial regime) \cite{Andrade1999}, PIV tracers face several problems. They may, for instance, damage the porous sample, and some of the flow regions may get excluded from the measurement due to the lack of tracers therein. Also, the tracers' size must be chosen carefully to reconcile the accuracy of the scattered signal acquisition and the minimal interaction between the tracers, flow, and the pore space \cite{SanchezGonzalez2018}. 
%	The remedy for this is to use numerical methods to solve fluid flow transport equations directly in pore space. Real samples, however, are relatively big compared to pore size, and using standard numerical methods based on computational grids may become impractical. This becomes most problematic at higher Reynolds numbers, where fine flow structures appear. 
	
	Modeling inertial flows numerically is of great importance to numerous branches of science and technology, e.g. CO2 storage, oil and shale gas extraction, electric battery design, or flow in circulatory system. It leverages many challenges faced in experimental studies such as Particle Image Velocimetry, for instance damaging the samples by the tracers or exclusion of some flow regions from the measurement due to the lack of the tracers therein.
	
	Lattice Boltzmann Method (LBM) is a popular numerical tool to solve the Navier-Stokes equations. It has been applied to the simulation of fluid transport in various contexts, e.g. porous media \cite{Matyka2011,Pan06,Gao15}, multiphase flows \cite{Falcucci2013}, semiclassical fluids flow \cite{Coelho2018}, particulate suspensions \cite{Ladd1994} and especially inertial flows,
	and turbulent flows \cite{Sun2022,Hegele2018}.
	Its advantages lie in the stable and local treatment of non-linear effects, a huge potential for parallelization of calculations, and simple implementation of basic no-slip boundary conditions on a regular grid \cite{Succi2018,Kruger2017,Guo2013}. Unfortunately, modeling inertial flows with LBM faces the challenge of memory and computational overhead. This is because, in LBM, increasing the Reynolds number means either a decrease in viscosity, an increase in the flow velocity, or an increase in the size of the system.	The first two approaches are limited by the simulation's stability. Thus, increasing the Reynolds number by increasing the size of the system is the only way to obtain stable, high-Reynolds number solutions in LBM. However, it comes with the cost of more lattice nodes inserted into the domain. Even the recently proposed LBM formulation with fixed relaxation time~\cite{Zhou2020,Matyka2021} loses its memory-consumption advantage at higher Reynolds numbers since one cannot tune the viscosity in this model.
	
	A possible aid may be to use irregular discretizations of space for the solution of the Lattice Boltzmann Equation (the so-called \textit{off-grid} LBM). It allows to discretize only the pore space and refine the discretization where it is needed. Especially, meshless LBM (MLBM)~\cite{Strzelczyk2024,Lin2019,Pribec2021,Musavi2019} has recently attracted attention because of the convenience and flexibility of use of the meshless approximation methods~\cite{Liu2005,Lancaster1981}. Recently we studied the interpolation-supplemented MLBM in low-Reynolds number flows~\cite{Strzelczyk2024,Strzelczyk24_2}. The decoupled space and velocity discretization of all off-grid LBM models makes it possible to increase the Reynolds number in the simulations simply by changing the streaming distance length relative to the system size, without inserting additional nodes into the domain. This procedure was first investigated in Lagrange interpolation-supplemented LBM~\cite{He97}.
	
	The aim of this work is to investigate the above-mentioned procedure of Reynolds number increase in MLBM. We benchmark the method in K\'arm\'an vortex street flow. Then, we compare its efficiency with the standard LBM to show the advantage achieved by the use of hybrid, locally refined discretizations. Further, we implement the meshless streaming step in the LBM1 model~\cite{Zhou2020,Matyka2021} to extend its applicability in higher Reynolds number flows. We test it in a flow through a porous sample in the Darcy and inertial regime and assess its memory savings compared to the lattice-based LBM with $\tau=1$.
	
	\section{Methods}
	\subsection{Lattice Boltzmann Method}
	
	Lattice Boltzmann Method \cite{Succi2018, Kruger2017} is a numerical tool for solving the discrete Boltzmann equation. During each timestep, two steps are performed - the streaming and the collision of the velocity distributions. The streaming is defined as
	\begin{equation}\label{eq:LBM_streaming}
		f_k(t+1,\boldsymbol{x}) = f_k^\text{post}(t,\boldsymbol{x}+\boldsymbol{e}_{k'})
	\end{equation}
	where $f_k$ is the $k$-th distribution function with its streaming vector $\bsym{e}_k$, $k'$ denotes the vector antiparallel to $\bsym{e}_k$ ($\boldsymbol{e}_k=-\boldsymbol{e}_{k'}$) and the superscript `post' denotes the post-collision distribution function. According to this notation, $\boldsymbol{x}+\boldsymbol{e}_{k'}$ denotes the departure node of the $k$-th population being streamed to $\boldsymbol{x}$. Eq. \eqref{eq:LBM_streaming} is written in a non-dimensionalized form with the timestep length and the lattice spacing equal to 1. 
	
	In this work, we use D2Q9 model with 9 streaming directions ($k=0,1,...,8$)
	\begin{equation}\label{eq:lattice_vectors}
		\boldsymbol{e}_k \in \left(
		[0,0],
		[1,0],
		[0,1],
		[-1,0],
		[0,-1],
		[1,1],
		[-1,1],
		[-1,-1],
		[1,-1]
		\right).
	\end{equation}
	Due to the Lagrangian nature of the streaming step, such a choice of the streaming vectors imposes the use of a square lattice (refer to Fig.~\ref{fig:lbm_vs_mlbm} for a graphical representation of the D2Q9 lattice and streaming directions). The collision step calculates $f^\text{post}_k$ from the current distributions and macroscopic fields. This work uses two relaxation time (TRT) collision term \cite{Ginzburg2008}
	\begin{equation}\label{eq:LBM_collision}
		f^\text{post}_k = f_k(t,\boldsymbol{x}) - \frac{1}{\tau^+}\left[f_k^+(t,\boldsymbol{x}) - f^{\text{eq}+}_k(t,\boldsymbol{x})\right] - \frac{1}{\tau^-}\left[f_k^-(t,\boldsymbol{x}) - f^{\text{eq}-}_k(t,\boldsymbol{x})\right]
	\end{equation}
	where $\tau^\pm$ are relaxation times for the symmetric and anti-symmetric parts of the distributions defined as
	\def\arraystretch{1.25}
	\begin{equation}\label{eq:trt_distributions}
		\begin{array}{c}
			\begin{aligned}
				f_k^+ = \frac{f_k+f_{k'}}{2}, &\quad& f_k^- = \frac{f_k-f_{k'}}{2} \\[.5ex]
				f_k^{\text{eq}+} = \frac{f_k^\text{eq}+f_{k'}^\text{eq}}{2}, &\quad& f_k^{\text{eq}-} = \frac{f_k^\text{eq}-f_{k'}^\text{eq}}{2}. \\
			\end{aligned}
		\end{array}
	\end{equation}
	The relaxation times are related to each other by
	\begin{equation}\label{eq:taus_relation}
		\Lambda = \left(\tau^+-\frac{1}{2}\right)\left(\tau^--\frac{1}{2}\right).
	\end{equation}
	For the non-fixed relaxation time models we use the value of $\Lambda=3/16$~\cite{Pan06}. The kinematic viscosity in the lattice units is
	\begin{equation}\label{eq:viscosity_lbm}
		\nu_{lb} = c_s^2\left(\tau^+-\frac{1}{2}\right)
	\end{equation}
	where $c_s=1/\sqrt{3}$ denotes the lattice speed of sound. The equilibrium distributions $f_k^\text{eq}$ are the functions of the local macroscopic density and velocity in lattice units, $\rho_{lb}\!=\!\rho_{lb}(t,\boldsymbol{x})$ and $\boldsymbol{u}_{lb}\!=\!\boldsymbol{u}_{lb}(t,\boldsymbol{x})$ respectively
	\begin{equation}\label{eq:feq}
		f^\text{eq}_k(t,\boldsymbol{x})
		\equiv
		f^\text{eq}_k(\rho(t,\bsym{x}),\bsym{u}_{lb}(t,\bsym{x}))
		=
		\omega_k\rho_{lb}
		\left[
			1
			+
			\frac{\boldsymbol{e}_k\cdot \boldsymbol{u}_{lb}}{c_s^2}
			+
			\frac{(\boldsymbol{e}_k\cdot \boldsymbol{u}_{lb})^2}{2c_s^4}
			-
			\frac{\boldsymbol{u}_{lb}^2}{2c_s^2}
		\right]
	\end{equation}
	where $\omega_k$ is the weight specific to the $k$-th streaming direction
	\begin{equation}
		\omega = \left(\frac{4}{9},\frac{1}{9},\frac{1}{9},\frac{1}{9},\frac{1}{9},\frac{1}{36},\frac{1}{36},\frac{1}{36},\frac{1}{36}\right).
	\end{equation}
	The macroscopic density and velocity in lattice units at time $t$ and point $\boldsymbol{x}$, are obtained from the discrete populations
	\begin{equation}\label{eq:macro_var}
		\renewcommand{\arraystretch}{2.5}
		\begin{aligned}
			\rho_{lb} =& \sum\limits_{k=0}^{q-1} f_k \\[0.5ex]
			\bsym{u}_{lb} =\>& \frac{1}{\rho_{lb}}\sum\limits_{k=0}^{q-1} f_k \bsym{e}_k
		\end{aligned}
	\end{equation}
	The pressure is related to the density via the lattice speed of sound: $p_{lb}=\rho_{lb}c_s^2$.
	
	In the numerical implementations, at each time step, Eq. ~\eqref{eq:LBM_collision} is first calculated to obtain the values of the post-collision distribution function (\textit{collision step}). Then, ~\eqref{eq:LBM_streaming} advects post-collision distributions to neighboring nodes (\textit{streaming step}). Because lattice nodes $\bsym{x}$ coincide with the departure/arrival nodes of the streaming step, transport is purely Lagrangian and amounts to an index shift in the distribution function array.
	
	The fixed-relaxation time variant of LBM (LBM1) assumes $\tau^+=1$ and $\Lambda=1/4$~\cite{Matyka2021,Zhou2020}, which by Eq.~\eqref{eq:taus_relation} sets $\tau^-=1$. In such case, Eq.\eqref{eq:trt_distributions} simplifies the streaming and collision procedures from Eqs.~\eqref{eq:LBM_streaming} and \eqref{eq:LBM_collision} to the same form as if Bhatnagar-Gross-Krook collision \cite{Bhatnagar1954} was used with the unit relaxation time:
	\begin{equation}\label{eq:LBM1_collision}
		f_k(t+1,\boldsymbol{x}) = f^\text{eq}_k(t,\boldsymbol{x}+\boldsymbol{e}_{k'}).
	\end{equation}
	As the equilibrium VDF, Eq.~\eqref{eq:feq}, at time $t+1$ will be fully defined by the fluid density and velocity, one can directly sum the $f_k^\text{eq}$'s in the above equation and obtain the evolution equations for the macroscopic moments, actually implemented in LBM1
	\begin{equation}\label{eq:LBM1_evolution}
		\begin{array}{rcl}
			\rho_{lb}(t+1,\bsym{x})
			&=&
			\displaystyle \sum\limits_{k=0}^{q-1} f^\text{eq}_k(\rho_{lb}(t,\boldsymbol{x}+\boldsymbol{e}_{k'}),\bsym{u}_{lb}(t,\boldsymbol{x}+\boldsymbol{e}_{k'}))\\
			\bsym{u}_{lb}(t+1,\bsym{x})
			&=&
			\dfrac{1}{\rho_{lb}}
			\displaystyle \sum\limits_{k=0}^{q-1} \bsym{e}_k f^\text{eq}_k(t,\boldsymbol{x}+\boldsymbol{e}_{k'}),\bsym{u}_{lb}(t,\boldsymbol{x}+\boldsymbol{e}_{k'})) \\
		\end{array}
	\end{equation}
	This substitution alleviates the need to store the distribution function explicitly. In each timestep, the equilibrium distributions are calculated directly from the current macroscopic fields according to Eq.~\eqref{eq:feq} and summed as in Eq.~\eqref{eq:macro_var} to obtain the macroscopic fields in the next time step.
	
	\subsection{Formulation of meshless LBM}
	
\begin{figure}[h!]
		\centering
		\includegraphics[height=.25\linewidth]{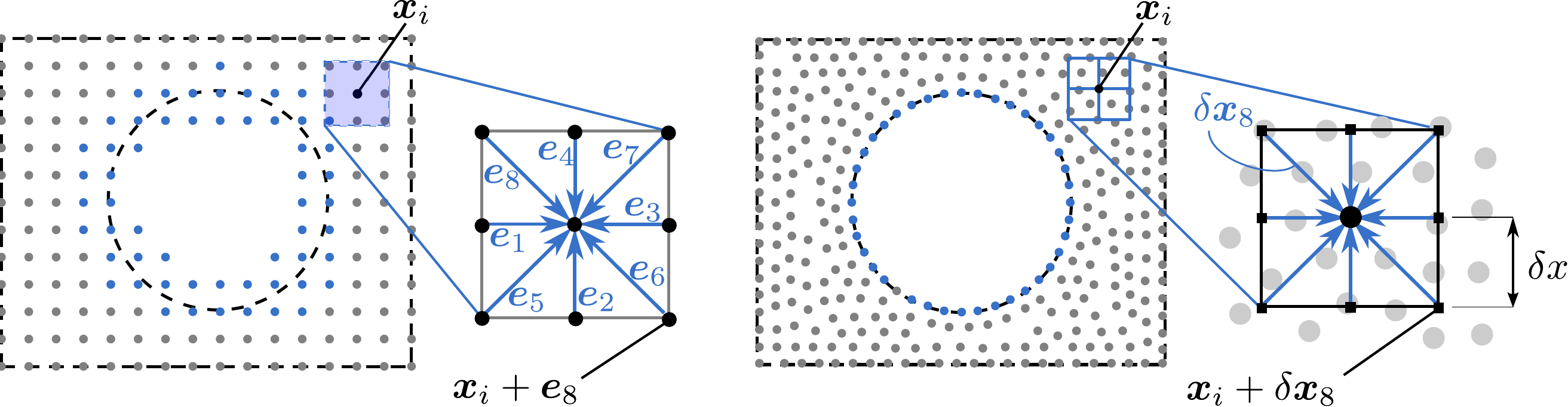}
		\caption{Comparison of space discretizations in the standard (\textit{left}) and meshless (\textit{right}) LBM. The standard LBM requires a square grid; the meshless LBM can operate on scattered, boundary-compliant node sets. Note that in MLBM, velocity is still discretized using a lattice, resulting in several departure points (so-called Lagrangian nodes, filled squares) assigned to each discretization point (so-called Eulerian nodes, filled circles). Source:~\cite{Strzelczyk2024}}
		\label{fig:lbm_vs_mlbm}
	\end{figure}

\begin{figure}[h!]
		\centering
		\includegraphics[height=.25\linewidth]{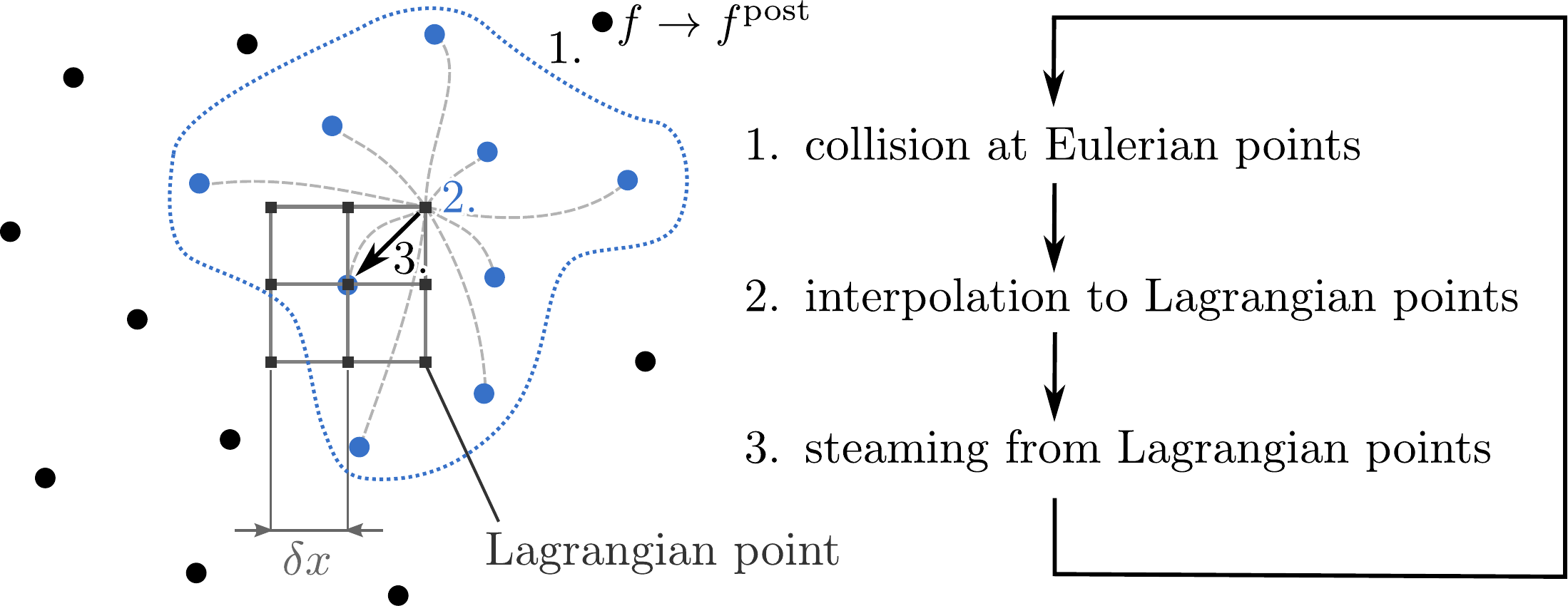}
		\caption{The procedure performed at a single timestep of the meshless LBM algorithm. Circles represent Eulerian points; squares denote Lagrangian points. A dashed loop encloses the stencil of a Lagrangian node, which consists of the nine closest Eulerian neighbors of this node (blue circles). The interpolated distribution function is streamed to the Eulerian point at the center of the presented square lattice. Source:~\cite{Strzelczyk2024}}
		\label{fig:olbm_basics}
	\end{figure}

	The meshless LBM algorithm considered in this work belongs to the family of off-grid LBM. Those methods trade the Lagrangian approach to solving the streaming step for the semi-Lagrangian~\cite{He1996,Strzelczyk2024,Kramer2017} or Eulerian~\cite{Pribec2021,Musavi2016} method, such that the space and velocity discretizations get decoupled, and nodes no longer need to be arranged in a regular grid. In the implementations taking the semi-Lagrangian approach, like the one discussed here, the discretization points $\bsym{x}$ (so-called Eulerian nodes) need not coincide with the departure points of the distribution function $\bsym{x}+\bsym{e}_{k'}$ (so-called Lagrangian nodes). Instead, with each Eulerian point we identify $q$ Lagrangian points to which the corresponding distributions are interpolated, and from which they are advected during the streaming step (see Fig.~\ref{fig:lbm_vs_mlbm}). We stress that the values of the distribution function in the Lagrangian points are not represented by an array in the computer memory, rather they are stored in temporary floating point variables during the iterations over the Eulerian points and then overwrite the previous-timestep values of $f_k$ in the Eulerian points' arrays. To relate the Eulerian and Lagrangian nodes' positions to one another, in MLBM the streaming vectors need to be expressed in physical units: $\delta\bsym{x}_k=\bsym{e}_k\delta x$ where $\delta x$ is the streaming distance. In this manner, the positions of the departure nodes of the $k$-th population are related to the positions of their Eulerian nodes as $\boldsymbol{x}+\delta\boldsymbol{x}_{k'}$. One introduces the interpolation step between the collision and the streaming to approximate the post-collision distributions in the appropriate Lagrangian points. Once the approximation is done, the streaming and the collision step are performed in the same way as in LBM (see. Fig.~\ref{fig:olbm_basics}). Thus, the streaming equation (counterpart of Eq.~\eqref{eq:LBM_streaming} in LBM) for MLBM becomes
	\begin{equation}
		f_k(t+1,\bsym{x}) = f_k^\text{post}(t,\bsym{x}+\delta\bsym{x}_{k'}) \approx \sum\limits_{i=1}^{N_L} f_k^\text{post}(t,\bsym{x}_i)w_i^k,
	\end{equation}
	where the sum goes over the nodes belonging to the interpolation stencil of the departure node $\bsym{x}+\delta\bsym{x}_{k'}$, $w_i^k$ are the interpolation coefficients (weights) for the $k$-th Lagrangian node, and the post-collision distributions $f_k^\text{post}$ are calculated the same way as in LBM, Eq.~\eqref{eq:LBM_collision}. The superscript $k$ at the interpolation weights stresses the fact that their value is different for each of $q-1$ departure nodes where the interpolation is performed. We note that in the models with zero-velocity population discretized (i.e. $\bsym{e}_0$ in Eq.~\eqref{eq:lattice_vectors}), one does not interpolate this population explicitly.
	
	Among many interpolation methods available to solve the streaming step on irregular discretizations, meshless methods \cite{Buhmann2017,Chen2014,Bhatia2016} are less computationally demanding and error-prone in the discretization process than the mesh-based schemes, they achieve boundary-compliant discretizations, offer high order approximations as well as feasible local refinement and adaptivity. Such approaches have been frequently used to solve transport partial differential equations \cite{Flyer2007,Shankar2015a,Slak2019}. The semi-Lagrangian streaming step in the discussed MLBM uses meshless interpolation based on radial basis functions-generated finite differences (RBF-FD)\cite{Buhmann2017}.	We choose cubic RBFs $\phi: \mathbb{R^D} \rightarrow \mathbb{R}, D=2$ with the centers $\bsym{x}_0$ at the approximation stencil members as the interpolation basis
	\begin{equation}\label{eq:cubic_rbf}
		\phi_0(\bsym{x}) = \phi(|\bsym{x}-\bsym{x}_0|) = \phi(r) = r^3.
	\end{equation}
	In this work, for RBF-FD approximation, we use stencils consisting of its $N_L=25$ closest Eulerian neighbors of a given Lagrangian node. To achieve high-order approximation, we use polynomial augmentation \cite{Flyer2016} of order two. For a detailed description of the algorithm, we refer the reader to our previous work \cite{Strzelczyk2024}.
	
	One of the drawbacks of RBF-FD approximation on scattered nodes is the higher number of stencil members needed to achieve the approximation of the same order and accuracy as, e.g. using standard finite differences on a regular lattice. In the context of this work, when the ratio $\delta x/h$ becomes very small, the $\sim \mathcal{O}(h^{p+1}/\delta x)$ error term typical for semi-Lagrangian schemes for advection can grow excessively, compromising the simulation's stability. Thus, an increase in the meshless approximation accuracy is needed, which inevitably leads to increasing either the number of Eularian nodes or the stencil size. To reduce the computational expense of our models for higher Reynolds numbers (small $\delta x/h$ ratios) we use hybrid scattered-regular discretizations. One part of the domain (usually the refined one, near the boundaries of complex shapes) is discretized with scattered points, while the rest of the domain (in the bulk of the fluid) uses regular, square discretization with a constant lattice parameter. For the Lagrangian points whose $9$ closest Eulerian neighbors do not form a regular lattice, we use RBF-FD interpolation, as described earlier. In other case, we use biquadratic Lagrange interpolation.
	
	The application of the above formalism to LBM1 is the following. The macroscopic density and velocities are interpolated to each of the Lagrangian points. Then, the equilibrium distributions are calculated in the Lagrangian nodes, Eq.\eqref{eq:feq}, and their values are summed according to Eq.\eqref{eq:macro_var} to obtain the next-timestep macroscopic fields in the Eulerian nodes. The interpolation of the macroscopic fields, rather than the equilibrium distributions, frees one from the need to store $f_k^\text{eq}$ in the memory. The evolution equation for the macroscopic moments (counterpart of Eq.~\eqref{eq:LBM1_evolution} in LBM1) can be thus stated for the meshless LBM1 (MLBM1) as follows
%	\begin{equation}\label{eq:mlbm1_streaming}
%		f_k(t+1,\bsym{x})
%		=
%		f_k^\text{eq}(\rho_{lb}(t,\bsym{x}+\delta\bsym{x}_{k'}),\bsym{u}_{lb}(t,\bsym{x}+\delta\bsym{x}_{k'})),
%	\end{equation}
%	where the interpolated macroscopic fields are
%	\renewcommand{\arraystretch}{2.5}
%	\begin{equation}\label{eq:mlbm1_interpolation}
%		\begin{aligned}
%			\rho_{lb}(t,\bsym{x}+\delta\bsym{x}_{k'})
%			&=&
%			\sum\limits_{i=1}^{N_L} \rho_{lb}(t,\bsym{x}_i)w_i\\
%			\bsym{u}_{lb}(t,\bsym{x}+\delta\bsym{x}_{k'})
%			&=&
%			\dfrac{1}{\rho_{lb}}\sum\limits_{i=1}^{N_L} \bsym{u}_{lb}(t,\bsym{x}_i)w_i\\
%		\end{aligned}
%	\end{equation}
%	\renewcommand{\arraystretch}{1.5}
	\renewcommand{\arraystretch}{2.5}
	\begin{equation}\label{eq:mlbm1_evolution}
		\begin{array}{rcl}
			\rho_{lb}(t+1,\bsym{x})
			&=&
			f^\text{eq}_0(\rho_{lb}(t,\bsym{x}),\bsym{u}_{lb}(t,\bsym{x}))
			+
			\displaystyle \sum\limits_{k=1}^{q-1} f^\text{eq}_k(\rho_{lb}(t,\bsym{x}+\delta\bsym{x}_{k'}),\bsym{u}_{lb}(t,\bsym{x}+\delta\bsym{x}_{k'}))
			\\
			\bsym{u}_{lb}(t+1,\bsym{x})
			&=&
			\dfrac{1}{\rho_{lb}}
			\displaystyle \sum\limits_{k=1}^{q-1}
				\bsym{e}_k f^\text{eq}_k(\rho_{lb}(t,\bsym{x}+\delta\bsym{x}_{k'}),\bsym{u}_{lb}(t,\bsym{x}+\delta\bsym{x}_{k'})),
			\\
		\end{array}
	\end{equation}
	\renewcommand{\arraystretch}{1.5}
	where we assume that the zeroth population is not interpolated, thus the summations do not concern it, and due to $\bsym{e}_0 = \bsym{0}$ it is omitted from the evolution of the macroscopic velocity. The macroscopic fields are interpolated to the departure points beforehand as
	\renewcommand{\arraystretch}{2.5}
	\begin{equation}\label{eq:mlbm1_interpolation}
			\begin{array}{rcl}
					\rho_{lb}(t,\bsym{x}+\delta\bsym{x}_k)
					&\approx&
					\displaystyle \sum\limits_{i=1}^{N_L} \rho_{lb}(t,\bsym{x}_i)w_i^k\\
					\bsym{u}_{lb}(t,\bsym{x}+\delta\bsym{x}_k)
					&\approx&
					\dfrac{1}{\rho_{lb}}
					\displaystyle\sum\limits_{i=1}^{N_L} \bsym{u}_{lb}(t,\bsym{x}_i)w_i^k.\\
				\end{array}
	\end{equation}

	A comparison of computational complexity of the discussed lattice-based and meshless LBM models is presented in Appendix~\ref{app:computational_expense}.
	
	\subsection{Increasing Reynolds number in meshless LBM}
	
	To assess the ratio between inertial and viscous forces in a flow, the dimensionless Reynolds number is used:
	\begin{equation}\label{eq:re_def}
		Re = \frac{UL}{\nu}
	\end{equation}
	arising naturally from non-dimensionalization of incompressible Navier-Stokes momentum equation:
	\renewcommand{\arraystretch}{2.5}
	\begin{equation}\label{eq:NSeq}
		\begin{array}{l}
			\dfrac{\partial \bsym{u}}{\partial t} + (\bsym{u}\cdot \nabla)\bsym{u}=-\dfrac{\nabla p}{\rho} + \bsym{g} + \nu \nabla^2\bsym{u} \\
			\dfrac{\partial \bsym{u^*}}{\partial t^*} + (\bsym{u^*}\cdot \nabla^*)\bsym{u^*}-\nabla^* p + \bsym{g}^* + \dfrac{1}{Re} \nabla^{*2}\bsym{u^*} \\
			\text{where } \bsym{x}^*=\bsym{x}/L, \quad \bsym{u}^* = \bsym{u}/U, \quad t^* = t\dfrac{U}{L}=tT, \quad p^* = p/\rho U^2, \quad \bsym{g}^* = \bsym{g}\dfrac{U^2}{L} \\
		\end{array}
	\end{equation}
	\renewcommand{\arraystretch}{1.5}
	where asterisk denotes non-dimensionalized quantities, $\nu$ is the fluid's kinematic viscosity, $\bsym{g}$ is the body acceleration, and $L$, $U$, $T$ denotes the reference length, velocity, and time, respectively. In LBM, the Reynolds number can be computed equivalently using the reference quantities in physical or lattice units. In the latter case, the reference velocity and viscosity are the model's parameters, while the reference length is the number of lattice links discretizing the chosen physical size, e.g. $9$ for the diameter of the cylinder in the left subplot of Fig.~\ref{fig:lbm_vs_mlbm}.
	
	Conventionally, when inertial forces dominate the system, the Reynolds number is much larger than unity, $Re \gg 1$. The practical utility of the Reynolds number lies in its ability to relate flows of different length and time scales to each other, e.g., flow past a real object versus its scaled model used in an experiment. The hydrodynamic coefficients like drag, lift, or permeability, are often considered as functions of $Re$, making the results more universal. Reynolds number also allows one to estimate the relative importance of non-linearity, turbulence, and momentum diffusion phenomena in a flow when choosing an appropriate physical model to describe the system.
	
\begin{figure}[h!]
		\centering
		\includegraphics[width=.35\linewidth]{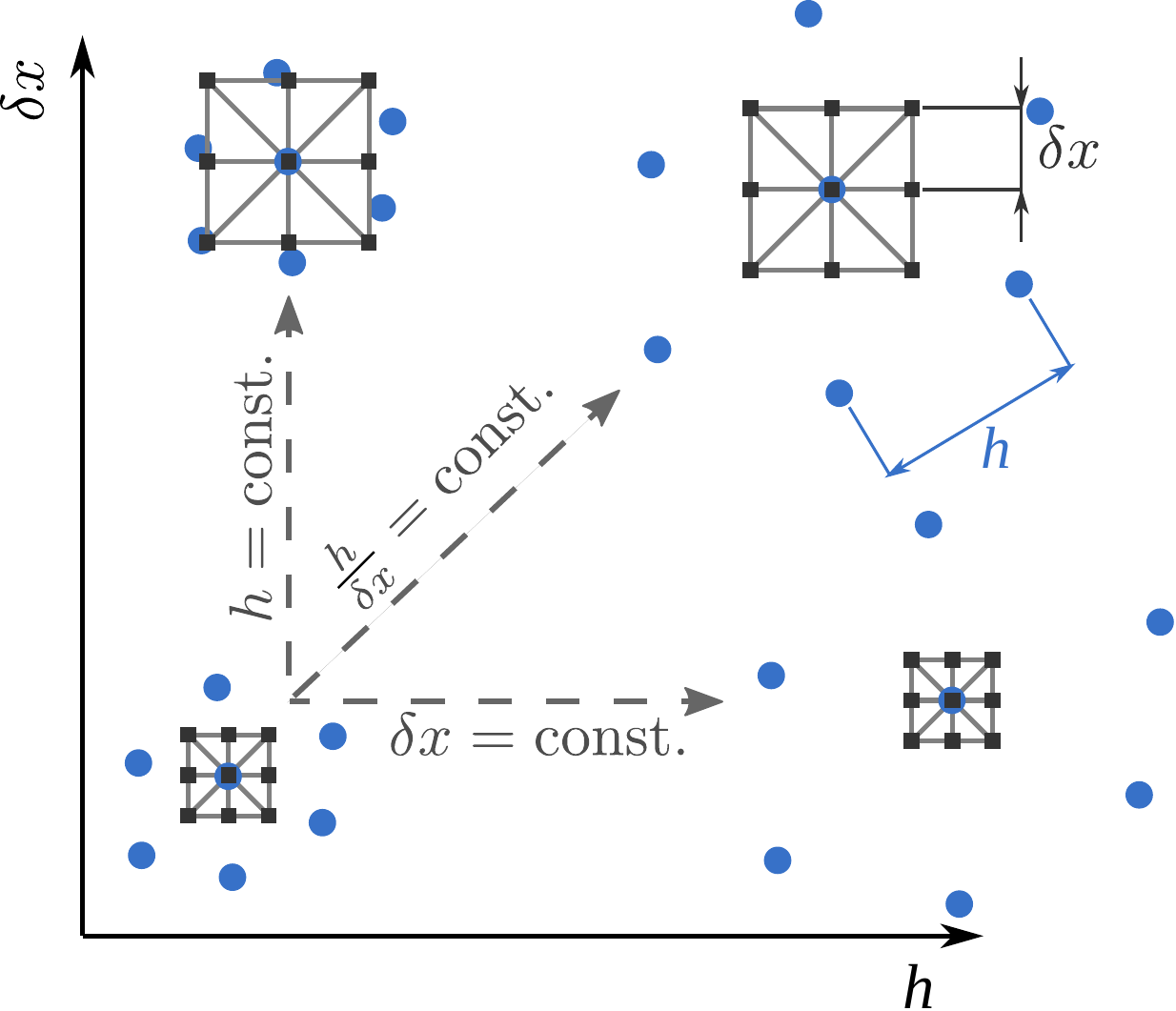}
		\caption{The $\delta x - h$ discretization parameters space and its three discussed directions (denoted by gray dashed arrows). Blue circles denote Eulerian points, and small dark gray squares denote Lagrangian nodes. Moving along each of the shown directions results in different scaling of the streaming length $\delta x$ and Eulerian discretization parameter $h$. In particular, in LBM, moving only along the constant $\delta x/h$ direction is possible.}
		\label{fig:dx-h_walk}
	\end{figure}

	The principle of the Reynolds number enhancement in MLBM (and off-grid LBM in general) we focus on in this work is as follows. In MLBM, the streaming distance length $\delta x$ can be varied independently from the positions of the nodes discretizing the domain. It means that the whole space spanned by the streaming distance $\delta x$ and the Eulerian discretization parameter $h$ is available in MLBM, in contrast to traveling along the $\delta x=h$ line in the standard LBM (see Fig.~\ref{fig:dx-h_walk}). For instance, consider a flow domain of size $L = 100$ in physical units. Let us assume that the streaming distance is $\delta x = 0.1$ (we do this by specifying the positions of Lagrangian points to differ from their Eulerian target points by $\delta\bsym{x}_k$ vectors) and that the inlet velocity and viscosity in lattice units remain fixed. With such a choice of $L$ and $\delta x$, the domain size in lattice units is $L_{lb}=L/\delta x = 1000$, regardless of $h$. Now, multiplying the streaming distance $\delta x$ by $\alpha_\delta < 1$ changes the domain size in lattice units $1/\alpha_\delta$ times. When the Reynolds number is calculated from the quantities in lattice units, the reference length $L_{lb}$ is larger than in the case of $\delta x=0.1$. At the same time, when $\delta x$ changes ($\delta x \rightarrow \alpha_\delta \delta x$), to keep the physical viscosity unchanged, timestep length must scale with the square of $\alpha_\delta$, since $\nu = \nu_{lb}\delta x^2/\delta t$. When this happens, the physical velocity $U$ changes into its counterpart in the scaled setup $U^\delta$ as
	\begin{equation}
		U = U_{lb} \frac{\delta x}{\delta t}
		\> \rightarrow \>
		U^\delta = U_{lb} \frac{\delta x \alpha_\delta }{ \delta t \alpha_\delta^2 }= U \> \frac{1}{\alpha_\delta}
	\end{equation}
	which in the case of decreasing the streaming distance ($\alpha_\delta<1$) means increasing the velocity $U$ in physical units.
	
	In general, the scaling of the streaming distance is followed by the scaling of the conversion factors from lattice units to physical units:
	\begin{equation}
		\begin{array}{lcrl}
			C_L = \delta x & \rightarrow & C_L^\delta = \delta x \alpha_\delta = &C_L\alpha_\delta \\
			C_t = \delta t & \rightarrow & C_t^\delta = \delta t \alpha_\delta^2 = &C_t\alpha_\delta^2 \\
			C_u = \delta x/\delta t &\rightarrow& C_u^\delta = \delta x \alpha_\delta / \delta t \alpha_\delta^2 = &C_u \alpha_\delta^{-1} \\
			C_\nu = \delta x^2 / \delta t &\rightarrow& C_\nu^\delta=&C_\nu\\
			C_g = \delta x / \delta t^2& \rightarrow& C_g^\delta = \delta x \alpha_\delta / \delta t^2 \alpha_\delta^4 = &C_g \alpha_\delta^{-3}.
		\end{array}
	\end{equation}
	
	Equivalently, one can think of scaling not the streaming distance by $\alpha_\delta$, but the positions of the Eulerian points by $\alpha_E = 1/\alpha_\delta$. Then, the reference length of the problem in physical and lattice units scales as $\alpha_E$. The reference velocity and viscosity remain intact both in lattice and physical units.
	
	We note that the discussed method of increasing the Reynolds number in MLBM should apply when Eulerian schemes are used for solving the streaming step~\cite{Lee2003}. The scaled streaming distance $\delta x$ would then be equivalent to the scaled streaming velocity in the term $\bsym{c} \cdot \nabla$ of the Boltzmann equation, where $\bsym{c}$ is the microscopic population velocity.
	
	\section{Results}\label{sec:results}
	
	\subsection{K\'{a}rm\'{a}n vortex street simulation using MLBM}\label{ssec:KVS}
	
\begin{figure}[ht!]
		\centering
		\includegraphics[height=.175\linewidth]{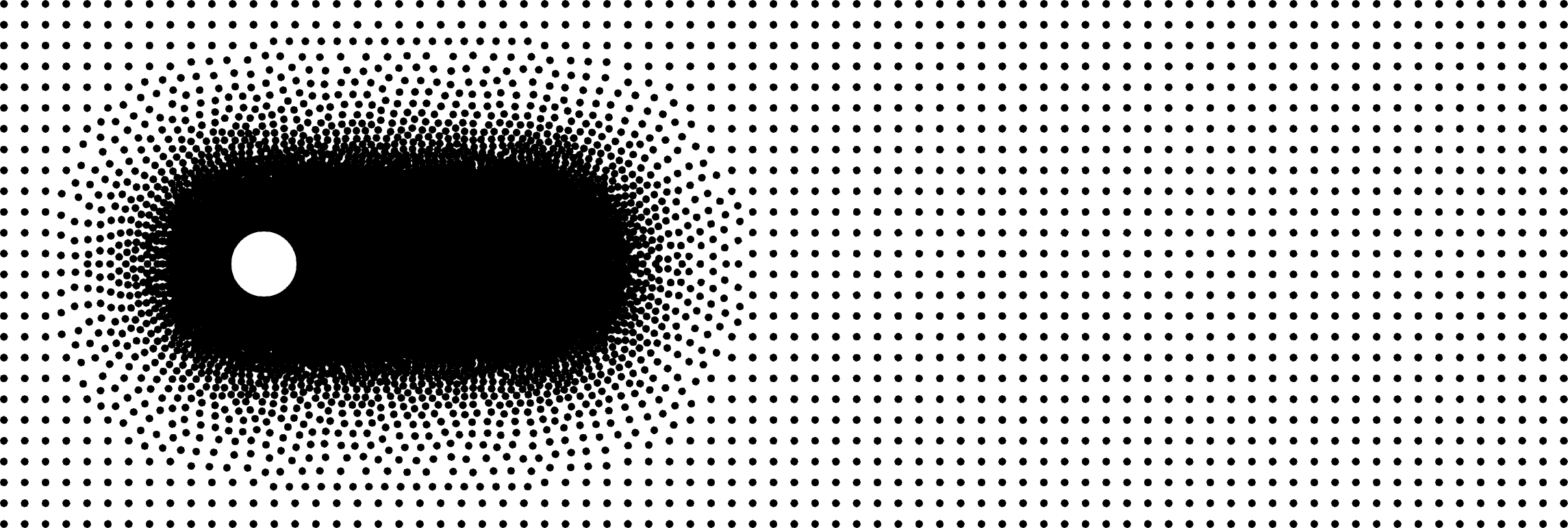}
		\hspace{1cm}
		\includegraphics[height=.175\linewidth]{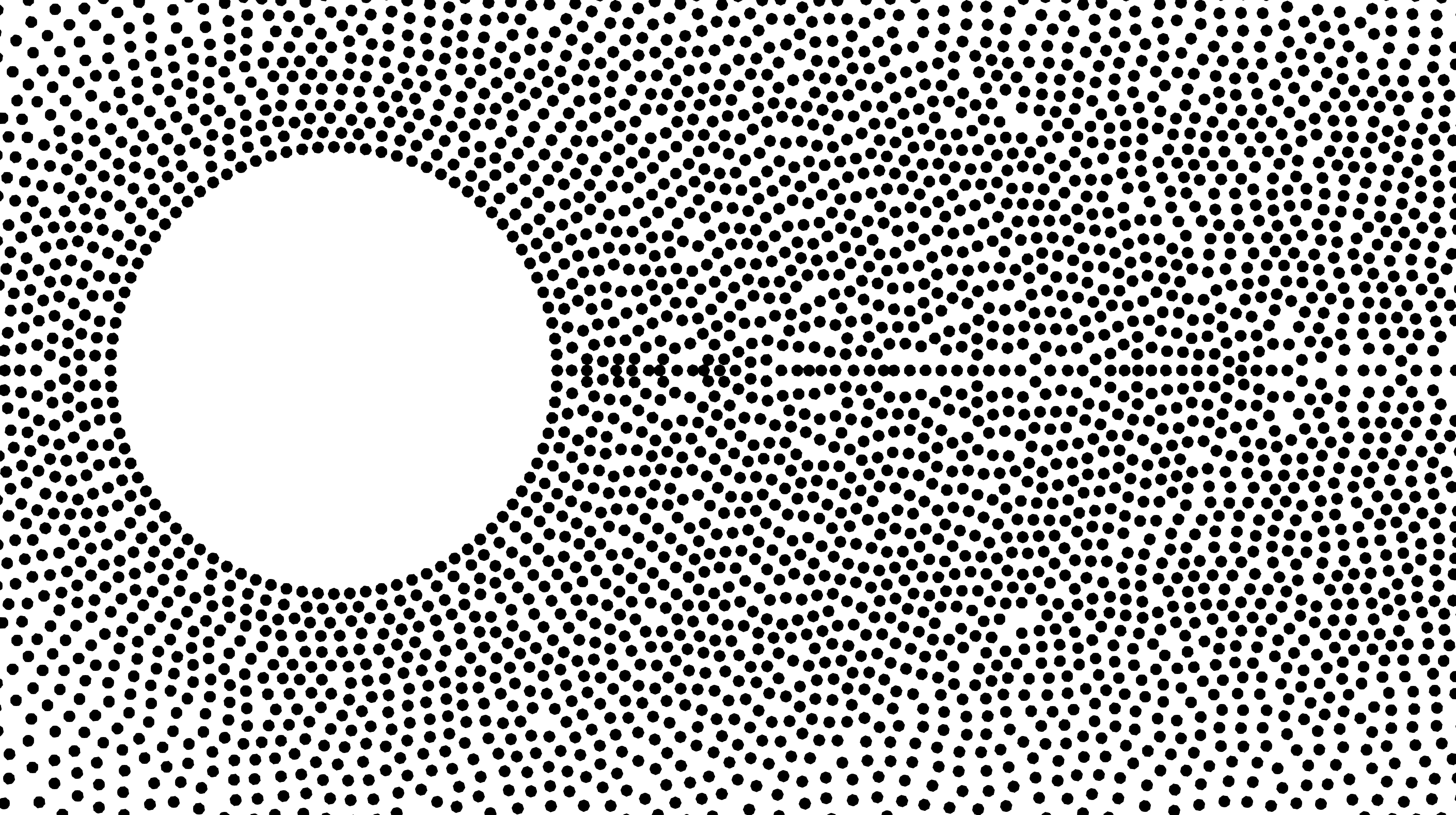}
		\caption{The hybrid regular-irregular discretization of the flow over a cylindrical obstacle used in the simulations (\textit{left}) and a zoom in at the discretization around the obstacle (\textit{right}).}
		\label{fig:CYLINDER_points}
	\end{figure}

	First, we investigate the flow over a cylindrical obstacle in a plane channel, Fig.~\ref{fig:CYLINDER_points}. The length of the channel is $3$, the height of the channel is $1$ and the cylindrical obstacle of diameter $d=0.14$ is placed on the channel's axis, at a distance $0.5$ from its inlet. The boundary conditions on the obstacle's surface is the no-slip wall. It is implemented using a simple bounceback rule for the populations for which the Lagrangian points lie inside the cylinder. The inlet and the outlet have the prescribed parabolic velocity profile with the maximum value of $U_{0,lb}=10^{-2}$ and unit density $\rho_{lb}=1$. Zero macroscopic velocity $\bsym{u}_{lb}=\bsym{0}$ and unit density $\rho_{lb}=1$ is imposed on the top and bottom boundaries. The four straight boundaries are implemented by overwriting all populations in their nodes with the equilibrium distributions parametrized with the desired macroscopic density and velocity. The initial condition is the equilibrium populations parametrized with the unit density $\rho=1$ and zero macroscopic velocity $\bsym{u}=\bsym{0}$.
	
	For MLBM simulations, we use the scattered discretization near the cylinder and in its wake and the regular discretization elsewhere. The distance between the nodes varies from $h_\text{min}=1/200$ on the cylinder's surface to $h_\text{max}=1/25$ away from the cylinder. The set of points used for the discretization is symmetric about the channel's axis to achieve a symmetric lift force profile in time. 
	The Reynolds number of each case was calculated based on the inlet velocity $U_0$ and the cylinder's diameter $d$:
	\begin{equation}
		Re = \frac{U_0 d}{\nu}.
	\end{equation}
	The values considered in MLBM simulations are $Re = 105,126,147,168$. We increase $Re$ by decreasing the streaming distance length from $\delta x = \pow{1.33}{-3}$ for $Re=105$ to $\delta x = \pow{8.33}{-4}$ for $Re=168$. The relaxation time is $\tau^+=0.53$ and the diffusive scaling of the timestep is used.
	
	For reference in timings analysis, we perform a series of standard LBM simulations with the same initial and boundary conditions as in the MLBM setup. The Reynolds numbers considered in LBM simulations are $Re=84,105,126,147,168$. The lattice parameter for $Re=84$ is $\delta x = 1/200$, and the increase in $Re$ is achieved by the appropriate decrease in $\delta x$, such that $Re=168$ yields the value $\delta x = 1/400$. We set the relaxation time to $\tau^+=0.51$, which is close to the stability limit. We use the diffusive scaling of the timestep. We note that MLBM setups use a higher value of $\tau^+$ due to stability issues.
	
	We investigate the onset of inertial effects in the flow qualitatively by observing the appearance of the K\'arm\'an vortex street in the obstacle's wake and quantitatively via the increase of the Strouhal number $St$:
	\begin{equation}
		St = \frac{fd}{U_0},
	\end{equation}
	where $f$ is the frequency of the wake vortices shedding.
	
	\begin{figure}[h!]
%		RESULTS FOR THE LBM COLORMAP ARE IN
%		/home/barman/zero_lokalnie/LBM/HOLMES_INERTIAL_test/res_LBM/2024-06-23_0/
		\centering
		\includegraphics[width=.75\linewidth]{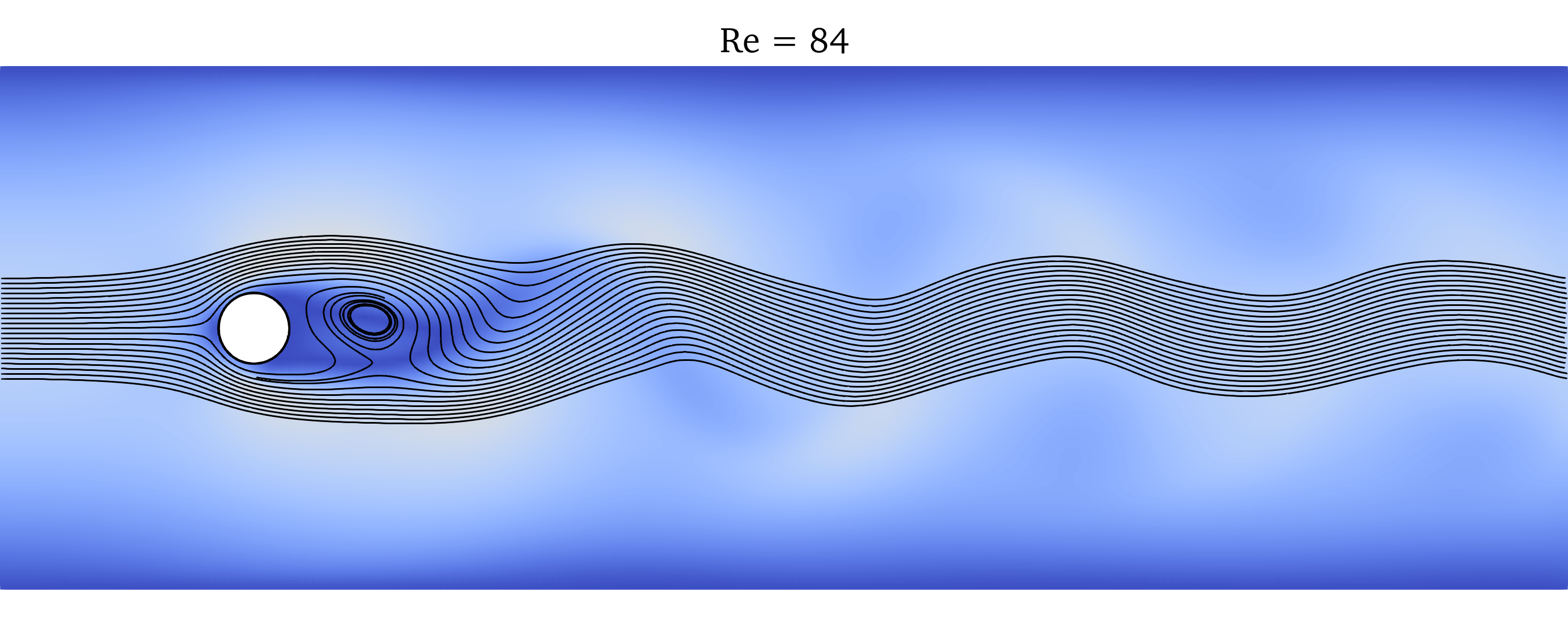}
		
		\vspace{.5cm}
		
		\includegraphics[width=.75\linewidth]{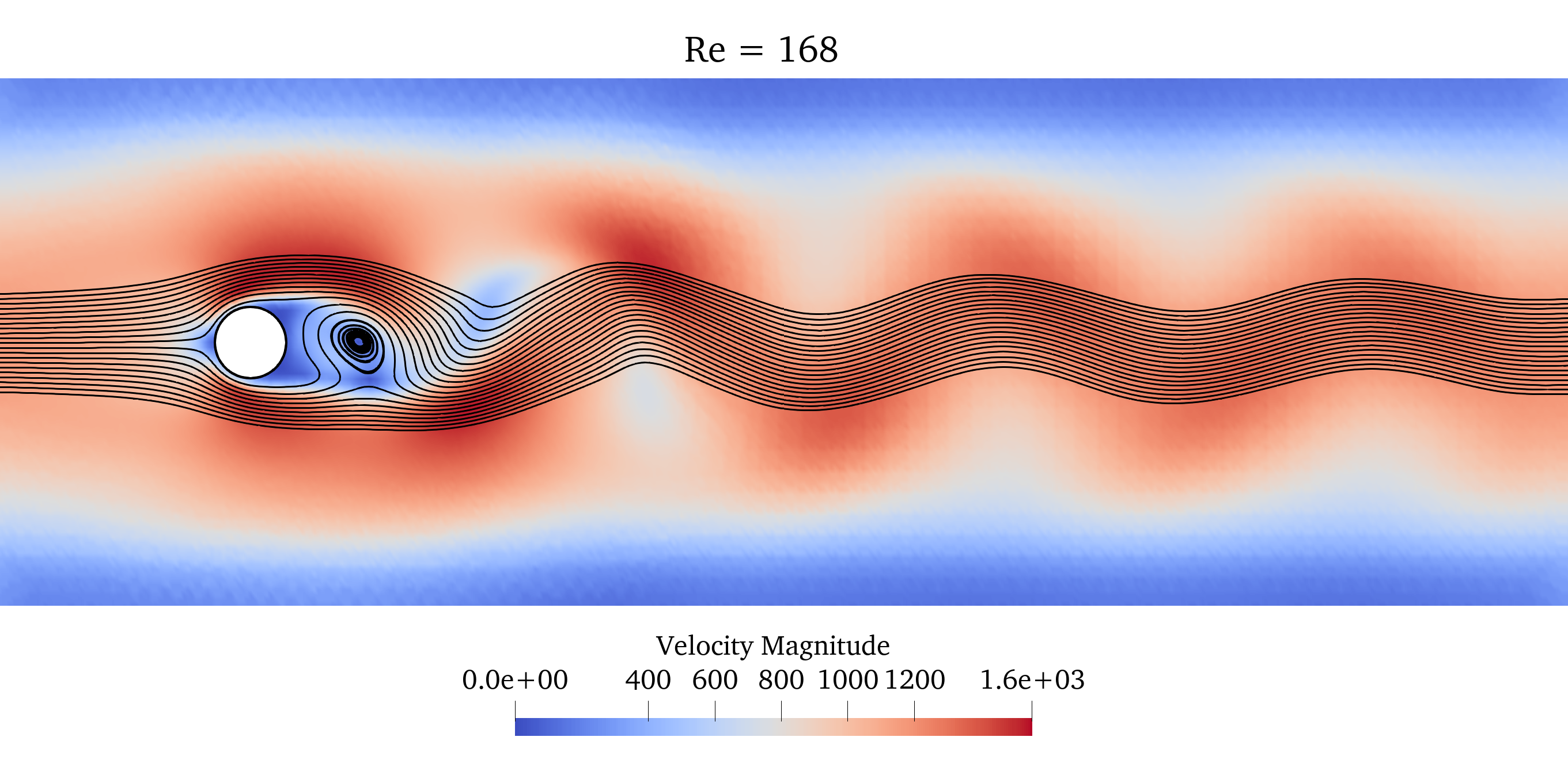}
		\caption{Streamlines of the velocity fields of the flow over a cylindrical obstacle for Reynolds numbers 84 (LBM results) and 168 (MLBM result). Both plots share the same color scale and the velocity magnitude is given in physical units.}
		\label{fig:DOMAINSCALE_channel}
	\end{figure}
	
	Fig.~\ref{fig:DOMAINSCALE_channel} shows velocity streamlines of the channel flow for two extreme Reynolds numbers: $Re=84$ obtained with LBM and $Re=168$ obtained with MLBM. The onset of inertial effects is visible as the distance between the repeating velocity field patterns in the obstacle's wake decreases.
	\begin{figure}[h!]
%		~/zero_lokalnie/LBM/MEDUSA_LBM_1/plot_forceVsTime.jl
		\centering
		\includegraphics[width=.5\linewidth]{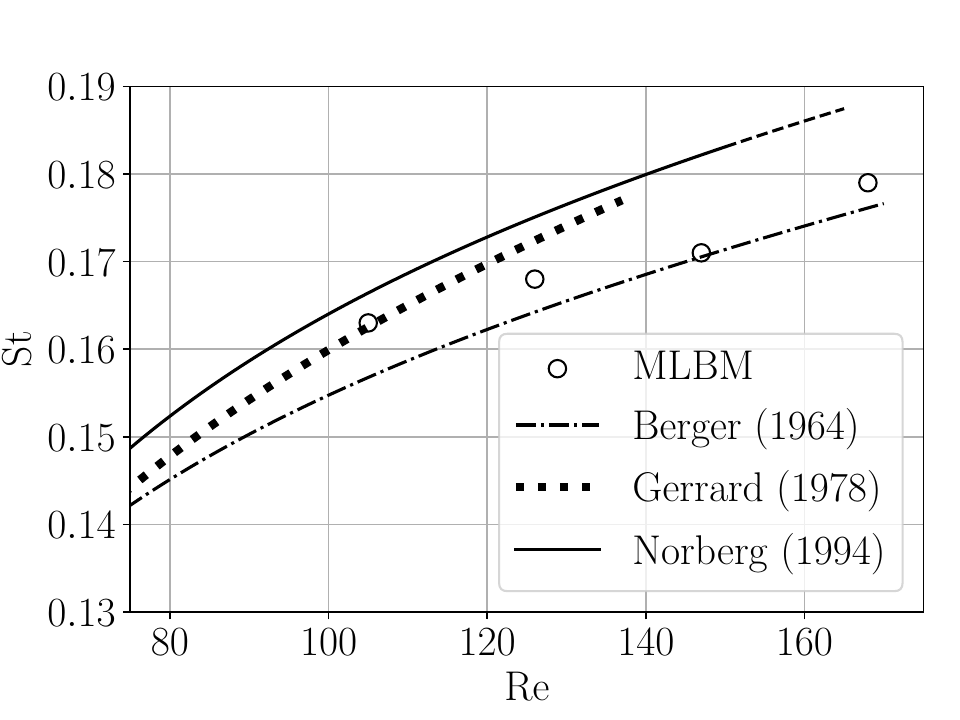}
		\caption{Strouhal number values achieved in flows around a cylindrical obstacle as a function of the flow's Reynolds number. The dashed part of the reference line for Norberg et al. \cite{Norberg1994} indicates the region where irregular shedding took place.}
		\label{fig:CYLINDER_Re_vs_St}
	\end{figure}
	Fig.~\ref{fig:CYLINDER_Re_vs_St} shows the values of the Strouhal number $St$ against the Reynolds number of the flows. The present results are in good agreement with a fit to the experimental data for the cylinder's diameter-to-length ratio equal to 2000 \cite{Norberg1994} and earlier experimental works \cite{Gerrard1978,Berger1964,Berger1964a}.
	\\
	
	\noindent \textit{Execution time analysis}
	
	\medskip
	
\begin{figure}[h!]
		\centering
		\includegraphics[height=.38\linewidth]{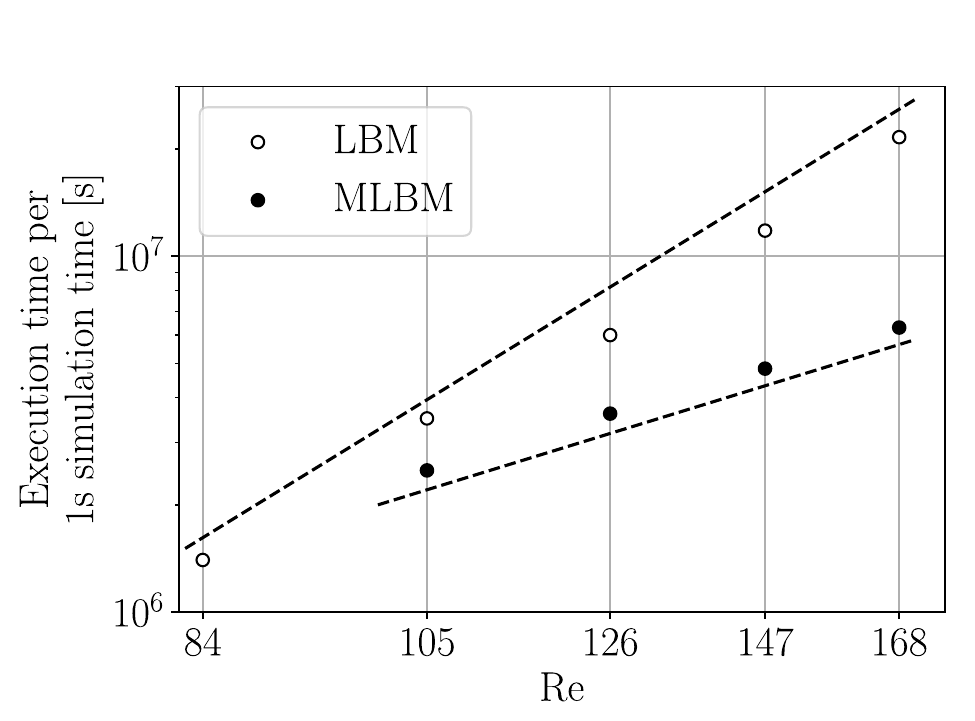}
		\caption{CPU time needed to execute 1s of the vortex shedding for LBM and MLBM in a single threaded run for various Reynolds numbers. Both axes use logarithmic scale. The dashed lines denote the 2nd and the 4th order slopes.}
		\label{fig:DOMAINSCALE_timings_per_1s}
	\end{figure}

	To compare the models in terms of their efficiency, LBM and MLBM codes are executed single-threaded on 2x Intel Xeon E5520 with 12GB RAM machine, and the execution times are measured using the C++ \texttt{<chrono>} header. Fig.~\ref{fig:DOMAINSCALE_timings_per_1s} shows the CPU time needed to simulate 1s of vortex shedding. For both methods, the execution time increases with $Re$. This is caused by the diffusive scaling of the timestep with the streaming distance $\delta x$ and the decreasing value of $\delta x$ with the increasing $Re$. This means that more timesteps are needed to reach 1s in the simulation for larger $Re$. At the same time, the execution time of MLBM rises slower than that of LBM ($\mathcal{O}(Re^2)$ vs. $\mathcal{O}(Re^4)$), since in the former, it is only the timestep number that increases with $Re$. For LBM, on the other hand, not only does the timestep number increase with $Re$, but also the total number of nodes in the domain.  For the highest value of $Re$ considered here, MLBM performs $\sim 3$ times faster than LBM.
	
\begin{figure}[h!]
		\centering
		\includegraphics[height=.4\linewidth]{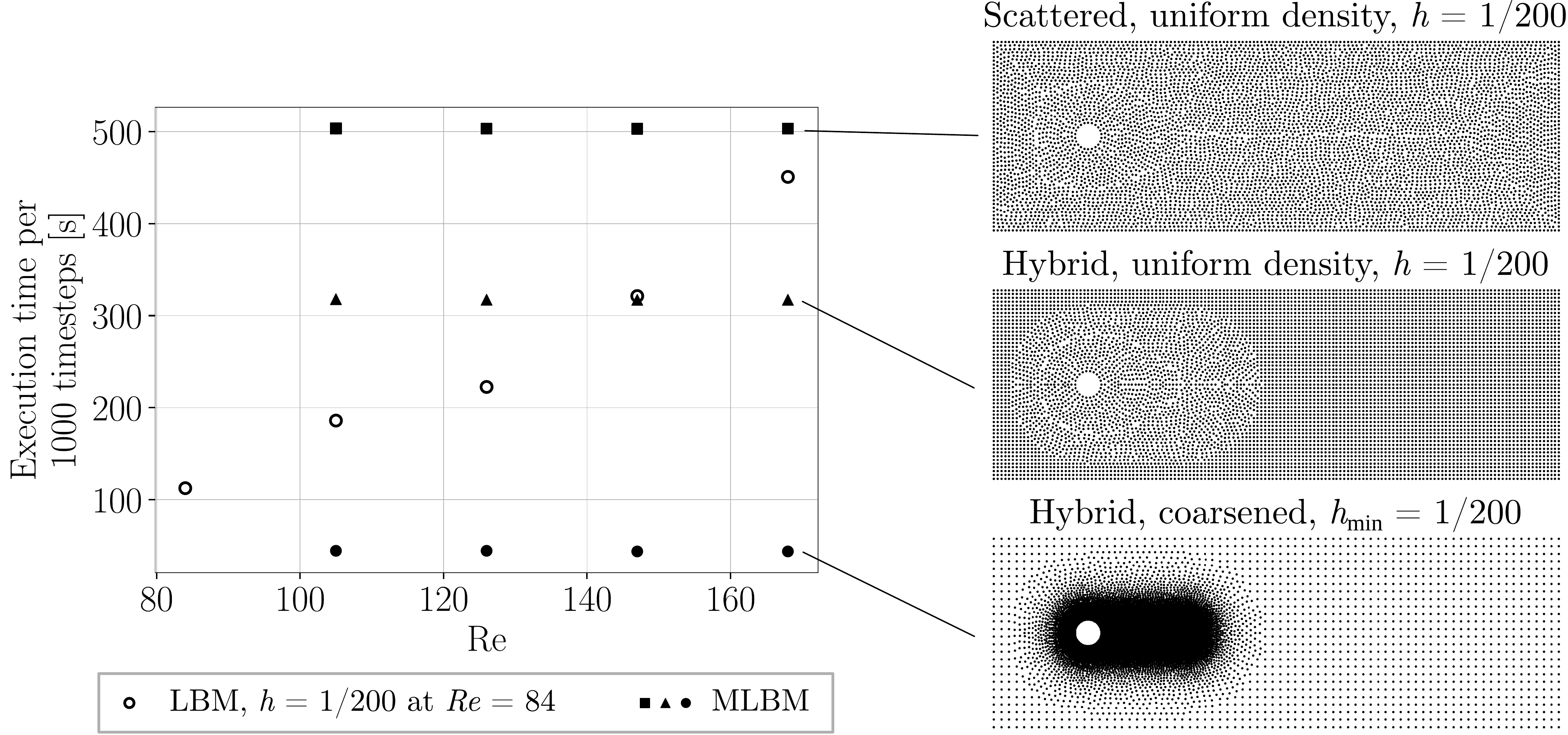}
		\caption{CPU time needed to execute $10^3$ timesteps of LBM and MLBM in a single threaded run for various Reynolds numbers. The visualizations of the discretizations of the three MLBM scenarios are shown on the right hand side. Note that two uniform visualizations show the discretizations coarser than $h=1/200$ for clarity.}
		\label{fig:DOMAINSCALE_timings_per_1kts}
	\end{figure}

	To further analyze MLBM speedup, we consider the CPU time needed to perform $10^3$ iterations in a single-threaded run on the same machine as before, shown in Fig.~\ref{fig:DOMAINSCALE_timings_per_1kts}. The data are presented for LBM and three scenarios of MLBM. The first scenario uses fully scattered discretization with RBF-FD approximation at each Lagrangian node, with uniform in space internodal distance $h=1/200$. In the second scenario, we exchange the scattered discretization away from the cylinder and its wake for a regular points arrangement. The third scenario additionally features coarser discretization away from the cylinder and is the one used in the calculation of the Strouhal number. The number of nodes in the LBM and MLBM setups is summarized in Table~\ref{tab:timings_nodecount}. The presented execution time increases with $Re$ for LBM because of the increasing number of nodes in the domain when the streaming distance $\delta x$ is decreased. This is in contrast to MLBM, where the increase of $Re$ is achieved only by decreasing the value of $\delta x$, with no additional nodes inserted in the domain. Due to this, the execution time stays constant for each scenario across the considered Reynolds numbers. Further, the execution times differ between the MLBM scenarios. About $40\%$ speedup between scenarios 1 and 2 is achieved by switching from RBF-FD approximation with $N_L=25$ nodes to the computationally cheaper biquadratic Lagrange interpolation away from the cylinder. It is achieved in spite of the increase of the total number of nodes by about $10\%$. Decreasing the discretization density away from the cylinder accounts for the subsequent $86\%$ speedup between scenarios 2 and 3.
	
	\begin{table}[h!]
		\caption{Number of nodes $N$ discretizing the domain in the LBM and MLBM simulations of the K\'arm\'an vortex street.}
		\label{tab:timings_nodecount}
		\centering
		\begin{tabular}[t]{lr}
			LBM case & $N$ \\
			\hline
			$Re=84$ & $119988$\\
			$Re=105$ & $187290$\\
			$Re=126$ & $269522$\\
			$Re=147$ & $366668$\\
			$Re=168$ & $478740$\\
		\end{tabular}
		\hspace{1cm}
		\begin{tabular}[t]{lr}
			MLBM case & $N$ \\
			\hline
			scattered uniform & $104825$\\
			hybrid uniform & $115185$\\
			hybrid coarsened & $10117$\\
		\end{tabular}
	\end{table}
	
	\subsection{Inertial effects in a flow through a porous sample using MLBM1}
	
\begin{figure}[h!]
		\centering
		\includegraphics[height=.35\linewidth]{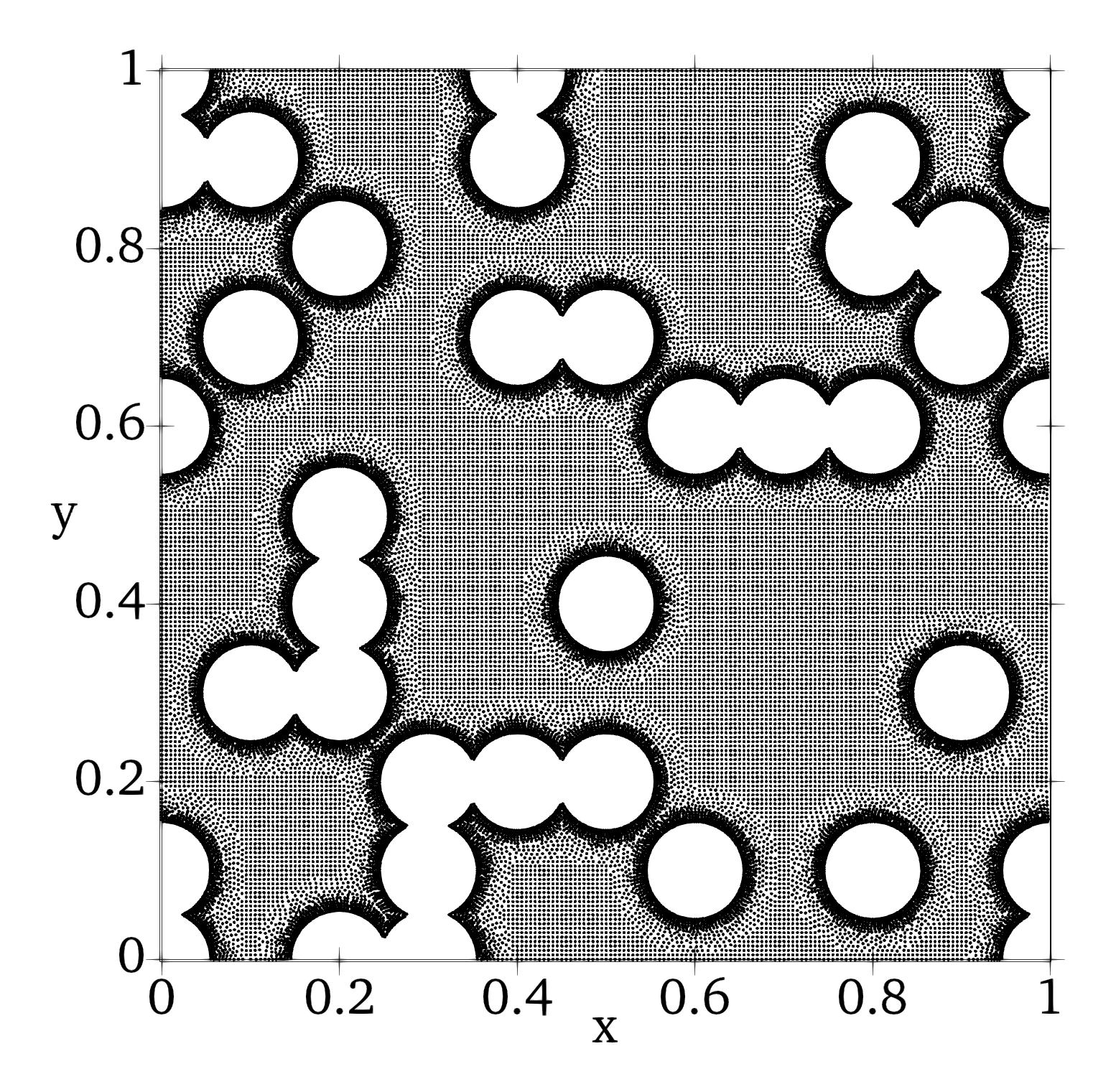}
		\hspace{1cm}
		\includegraphics[height=.35\linewidth]{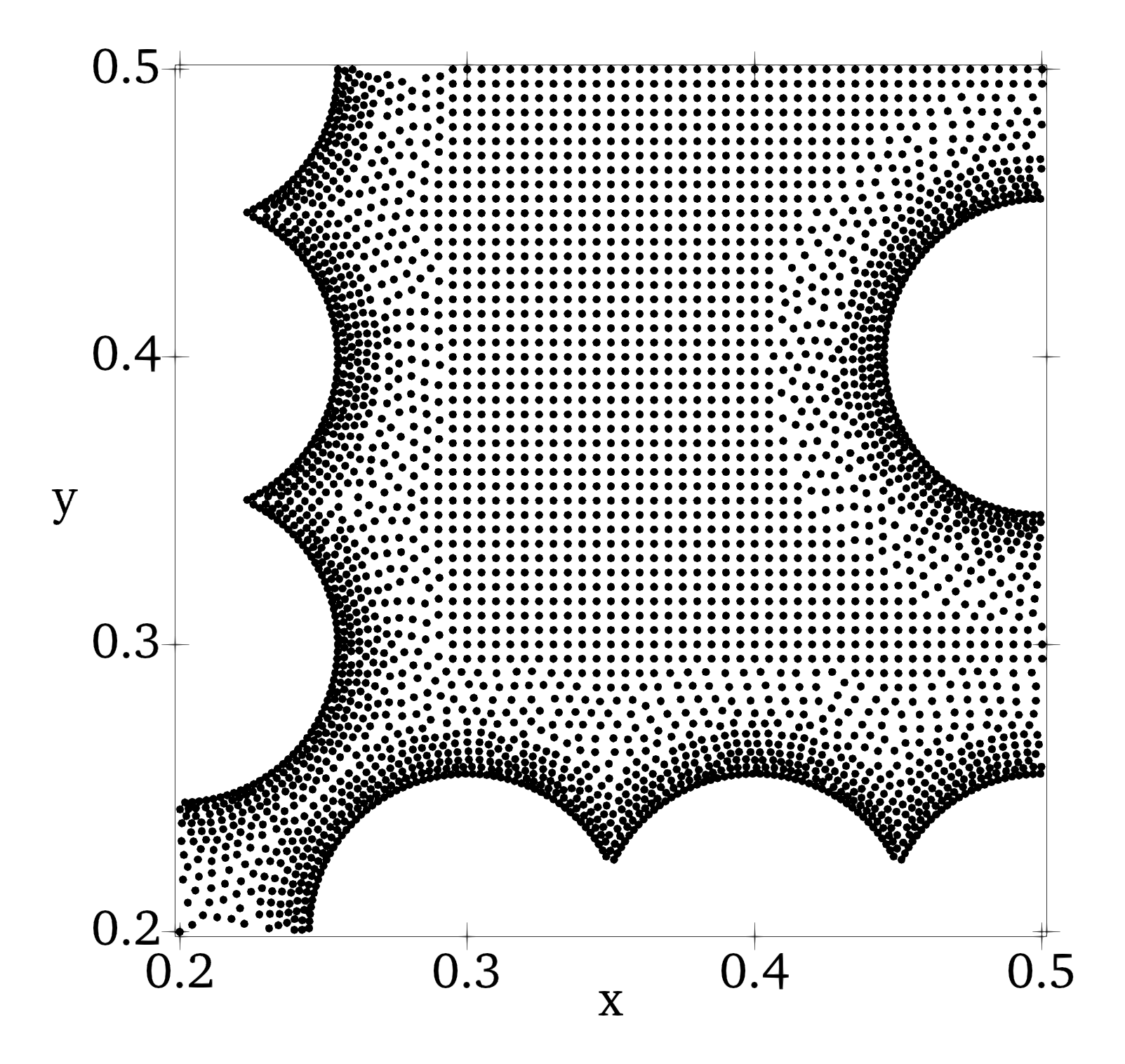}
		\caption{The point cloud used for the MLBM1 simulations of a flow through a porous sample: the whole discretized domain and a zoom at the discretization detail (\textit{right}).}
		\label{fig:POROUS_points}
	\end{figure}

	We investigate the onset of inertial effects in a flow through a porous medium of porosity $\varphi=0.64$ with MLBM1. We use hybrid space discretization (see Fig.~\ref{fig:POROUS_points}) -- irregular points arrangement within distance $0.04$ from the obstacles and regular square arrangement elsewhere. The internodal distance is $h=5\cdot 10^{-3}$ in the bulk of fluid and it linearly decreases over the scattered nodes volume to $h=2.5\cdot 10^{-3}$ on the obstacles' surface. The no-slip boundary condition is imposed on the top and bottom boundary and the circular obstacles' surfaces using the interpolated bounce back rule~\cite{Pan06}. The boundary Eulerian points are placed exactly on the no-slip walls, which simplifies the interpolated bounce back for the unknown $k$-th population to
	\begin{equation}
		f_k(t+1,\bsym{x}) = f_{k'}^\text{post}(t,\bsym{x}+\delta \bsym{x}_k)
	\end{equation}
	The periodic boundary condition is imposed on the left and right boundary by periodic search of the stencil members. The flow is forced by a constant acceleration of value $g_{lb}=10^{-7}$ along the $x$ direction with the second-order discretization in velocities. The investigated values of the streaming distance $\delta x$ range from $2 \cdot 10^{-3}$ to $1.5 \cdot 10^{-4}$. The timestep length is chosen to give the value of the kinematic viscosity in physical units $\nu=1$ and takes the values from the range $\delta t \in [6.67 \cdot 10^{-7} ; 3.75 \cdot 10^{-9}]$. The Reynolds number for each case is calculated based on the mean flow velocity in the $x$-direction at timestep $n$
	\begin{equation}\label{eq:mean_u_velocity}
		\langle u \rangle^n = \frac{1}{\phi}\int\limits_D d\Omega \> u(n\delta t,\bsym{x}),
	\end{equation}
	sample's side length equal to $1$ and kinematic viscosity $\nu$. The steady state is assumed to be achieved when the relative difference of the mean $x$-component of velocity between $\Delta n_t = 2 \cdot 10^4$ consecutive timesteps defined as
	\begin{equation}
		\Delta u
		=
		\dfrac{1}{\langle u \rangle^n}
		\frac{
			\langle u \rangle^{n+\Delta n_t}
			-
			\langle u \rangle^n}
		{\delta t \Delta n_t}
	\end{equation}
	falls below $10^{-2}$. For the systems that eventually become unsteady, namely $\delta x > 2.1\cdot 10^{-4}$,  we run the simulations up to $1.56 \cdot 10^6$ timesteps. In the subsequent analysis, the discussed velocity fields and hydrodynamic parameters will be those taken from the last timestep of the simulations. We validate the MLBM results by performing the standard (non-fixed relaxation time) LBM simulations with lattice parameter $\delta x = 2.5 \cdot 10^{-3}$, simple bounceback, $g_{lb} = 10^{-7}$, and the value of $\tau^+$ chosen such that the desired acceleration in physical units $g=g_{lb} \> \delta x / \delta t^2$ is achieved.
	
\begin{figure}[h!]
		\includegraphics[height=.33\linewidth]{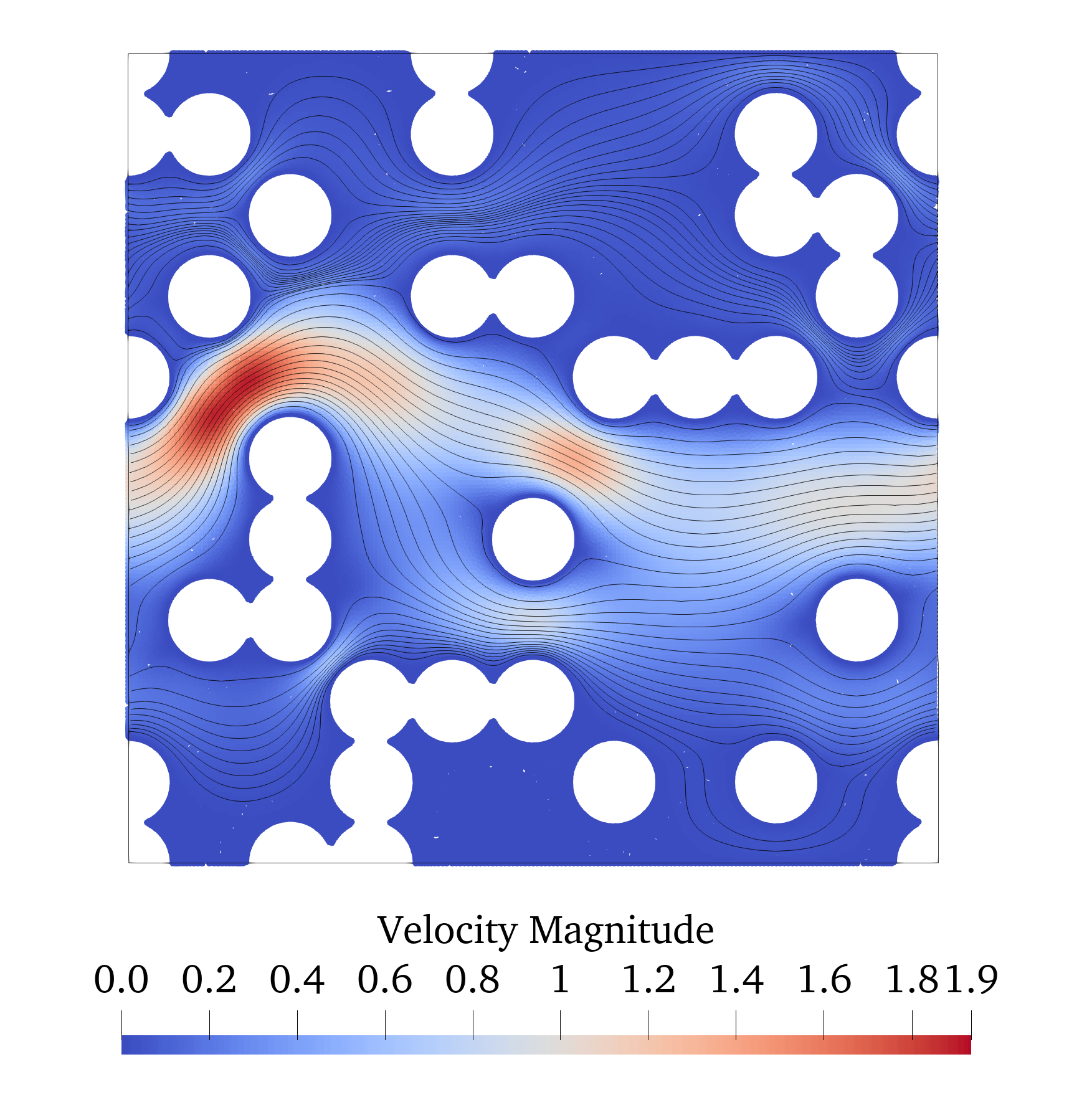}
		\includegraphics[height=.33\linewidth]{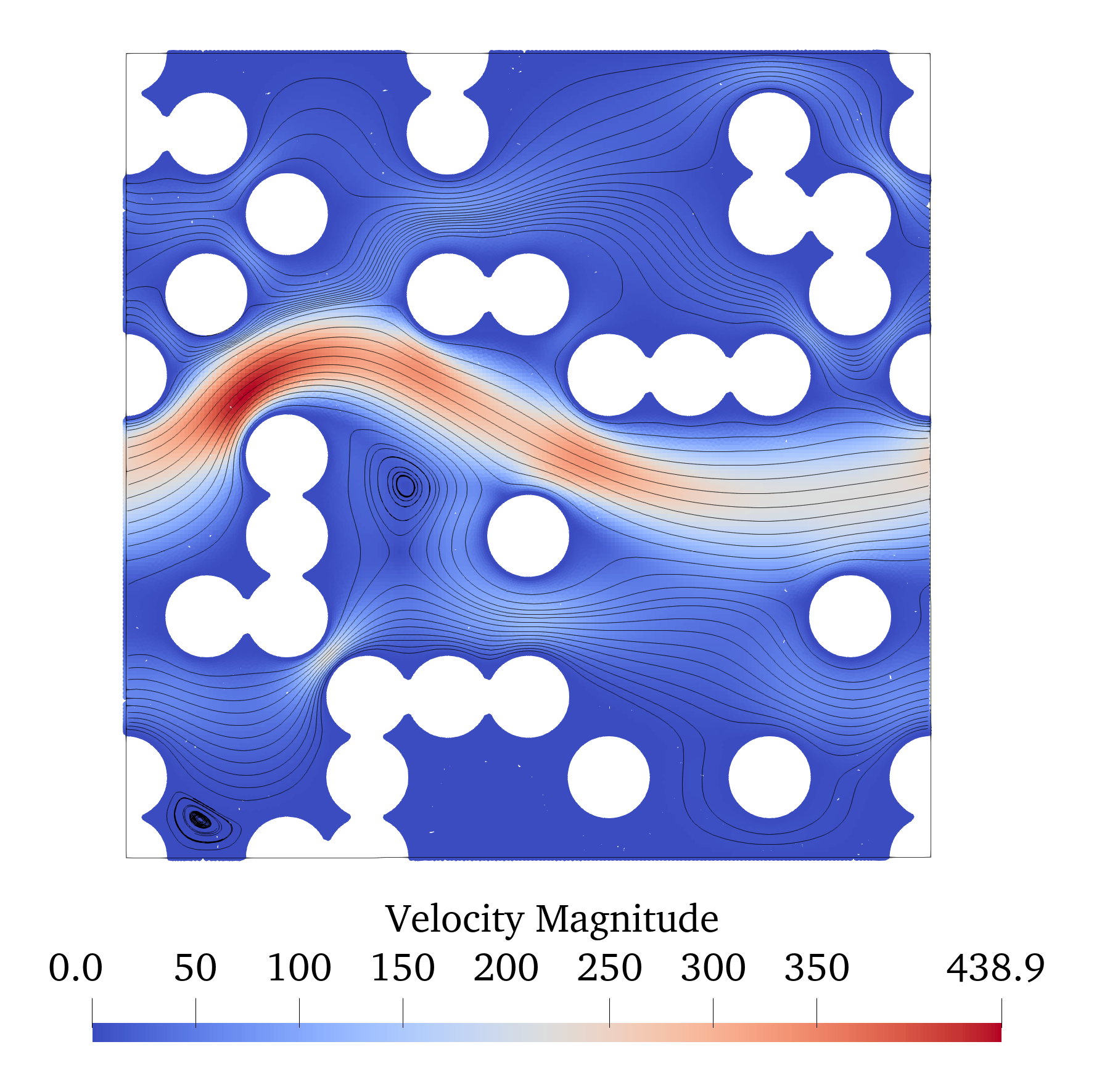}
		\includegraphics[height=.33\linewidth]{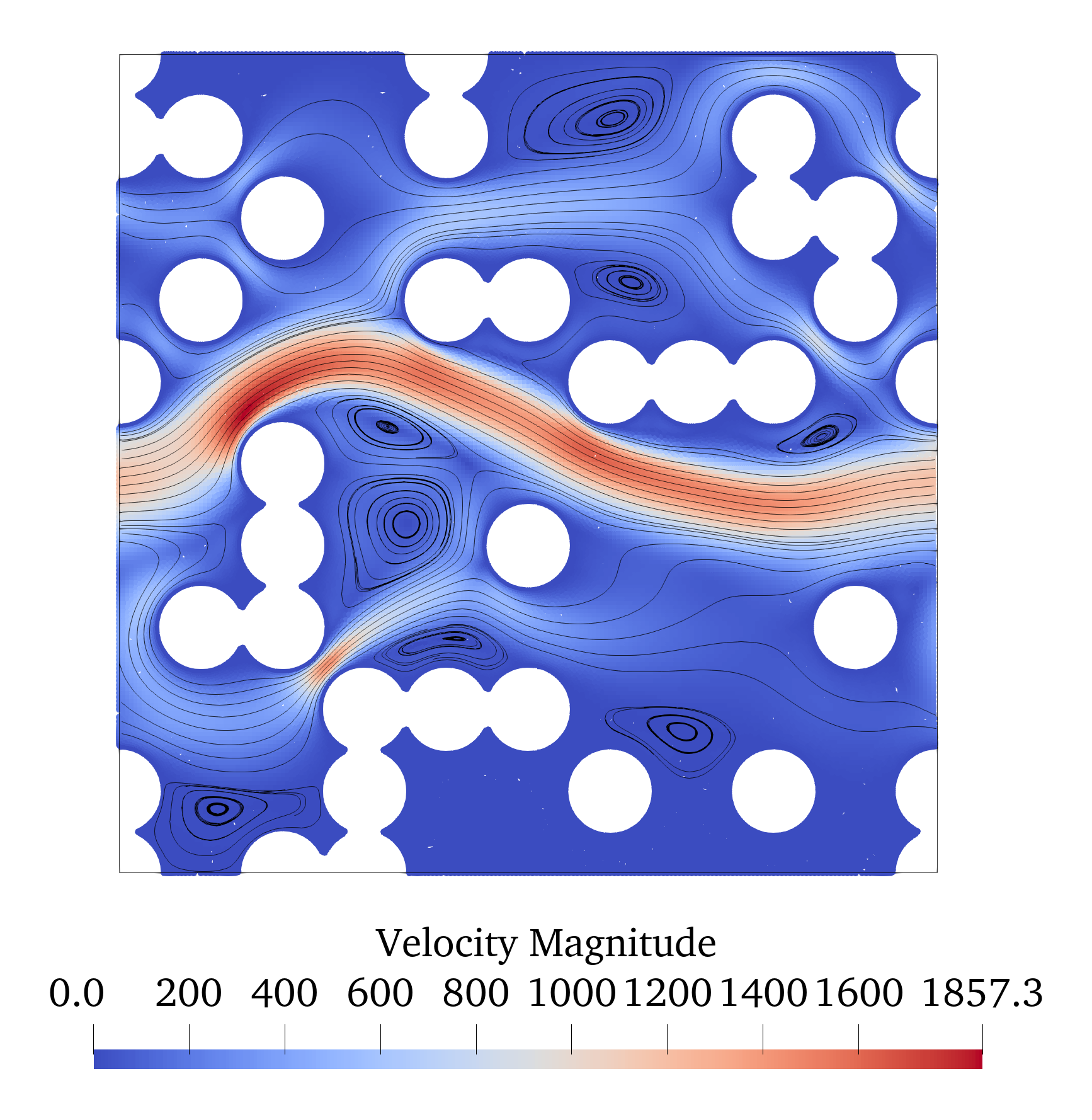}
		\caption{The results of flows through the porous domain obtained with MLBM1: streamlines and magnitude colormaps of the final timestep velocity field. The streaming distances are $\delta x = 2\cdot 10^{-3}$, $3\cdot 10^{-4}$, and $1.5\cdot 10^{-4}$ ($Re=0.26$, $60.73$, and $262.84$, respectively, from left to right).}
		\label{fig:DOMAINSCALE_porous}
	\end{figure}

	Fig.~\ref{fig:DOMAINSCALE_porous} shows the visualization of the velocity field for three chosen values of $\delta x$: in the steady state for $2\cdot 10^{-3}$ and $3\cdot 10^{-4}$, and at the final iteration for $1.5\cdot 10^{-4}$. As the Reynolds number increases, new recirculation zones appear, and the existing ones grow, causing the streamlines to separate from the grains' wakes.
		
\begin{figure}[h!]
		\centering
		\includegraphics[height=.36\linewidth]{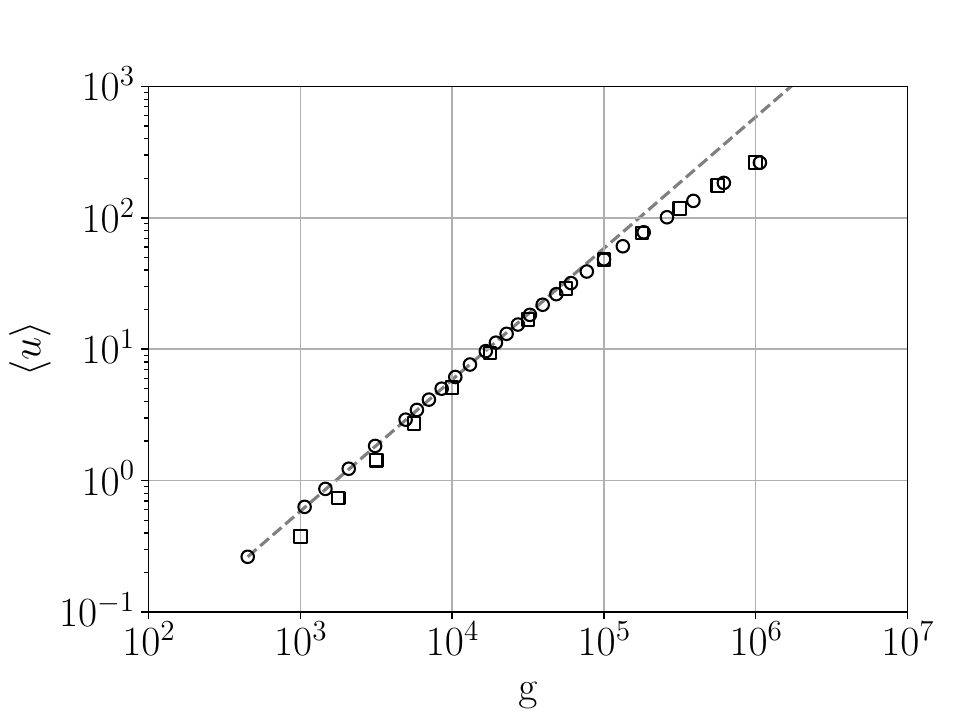}\hspace{.05cm} 
		\includegraphics[height=.36\linewidth]{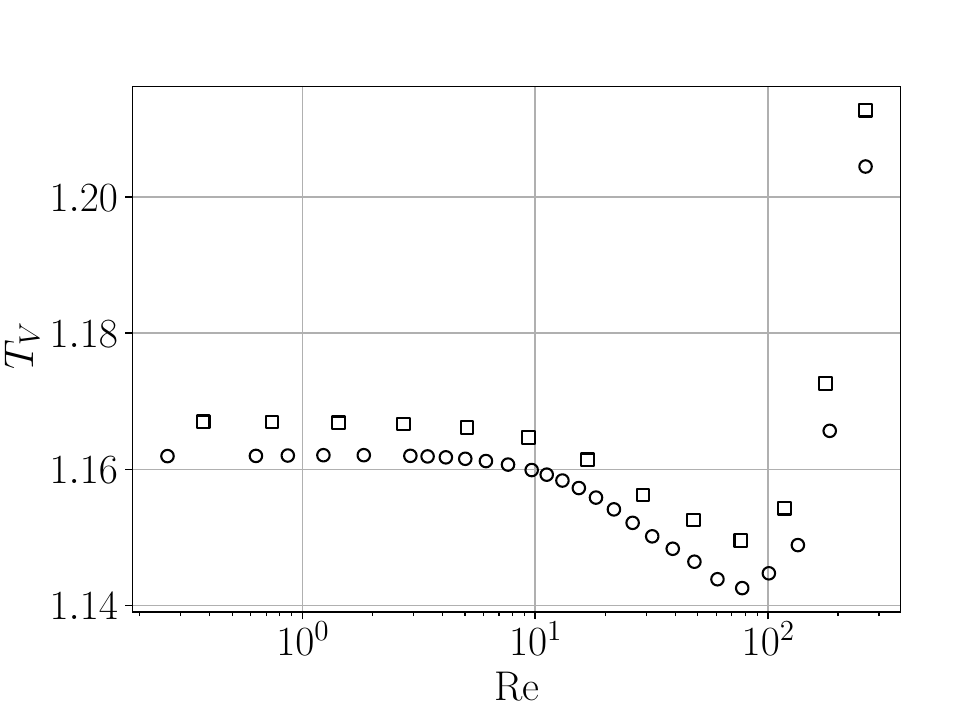}
		\caption{Quantitative measures of the inertial effects for the flows through the porous sample obtained with the concerned MLBM model (\textit{circles}) and validation with the standard LBM (\textit{squares}). \textit{Left:} mean $x$-component velocity defined by Eq.~\eqref{eq:mean_u_velocity} as a function of the acceleration $g$ forcing the flow. The dashed line denotes the 1st order slope. \textit{Right:} tortuosity $T_V$ defined by Eq.~\eqref{eq:tortuosity} versus Reynolds number.}
		\label{fig:POROUS_umean_vs_alpha}
	\end{figure}

	The appearance of inertial effects can be quantified in terms of the deviation from Darcy's law~\cite{Whitaker1986,Neumann1977}. The left subplot of Fig.~\ref{fig:POROUS_umean_vs_alpha} shows the dependence of the mean $x$-component of the velocity $\langle u \rangle$ on the acceleration $g$ forcing the flow. For $g \ge 4 \cdot 10^4$ ($Re = \langle u \rangle = 20$), the relation $\langle u \rangle \propto g$ changes to a sublinear one, which is connected to the increased energy dissipation in the vortices. Another indicator of the onset of inertia in the system is the tortuosity $T_V$~\cite{Matyka2011,Duda2011} calculated with the integrals of the velocity field
	\begin{equation}\label{eq:tortuosity}
		T_V = \dfrac{\langle |\bsym{u}| \rangle}{\langle u \rangle},
	\end{equation}
	where the triangular brackets have the same meaning as in Eq.~\eqref{eq:mean_u_velocity} and we omitted the notion of the timestep for simplicity. Its dependence on the Reynolds number $Re$ is shown in the right subplot of Fig.~\ref{fig:POROUS_umean_vs_alpha}. After the initial plateau, the tortuosity falls down in the range $Re \in [3; 60]$ to later rapidly increase. The range of $Re$ where $T_V$ decreases includes the approximate transition Reynolds number obtained from Darcy's law analysis from the left subplot. A similar behavior of tortuosity for steady-state systems is observed in our previous work~\cite{Sniezek2024}. For both $\langle u \rangle$ and $T_V$, the values obtained with MLBM exhibit satisfactory compliance with those from LBM.

	\medskip
	
	\noindent \textit{Computational complexity and memory demands of the MLBM1 algorithm}
	
	The memory demands of LBM1 yield
	\begin{equation}\label{eq:lbm1_mem}
		M_\text{LBM1} = N_\text{LBM1}\cdot2N_fc
	\end{equation}
	where $N_\text{LBM1}$ is the number of nodes in the discretization, $N_f$ is the number of macroscopic fields (in our case 3 -- $\rho$,$u$ and $v$), and $c$ is the number of bytes per single number (e.g. 8 for \texttt{double})~\cite{Matyka2021}. Compared to this, MLBM1 uses additional memory to store the arrays of interpolation weights and stencil members indices for each of the $N(q-1)$ Lagrangian nodes (the term $-1$ appears because no interpolation is needed for the populations with zero microscopic velocity). In the present work, each RBF-FD interpolation stencil consists of $25$ nodes, and the Lagrange interpolation stencils need at most $9$ stencil members, thus one can express the memory demands for MLBM1 as
	\begin{equation}\label{eq:mlbm1_mem}
		M_\text{MLBM1} = \left[\underbrace{2 N_f N_\text{MLBM1}}_\text{macroscopic fields}
		+
		\underbrace{\dfrac{3}{2}(25 \cdot N_\text{MLBM1,RBF-FD} + 9 \cdot N_\text{MLBM1,Lag})(q-1)}_\text{inteprolation weights and stencil indices}\right]c
	\end{equation}
	where $N_\text{MLBM1,RBF-FD}$ and $N_\text{MLBM1,Lag}$ denote the number of Eulerian nodes having irregular and regular stencils, respectively, and  $N_\text{MLBM1}=N_\text{MLBM1,RBF-FD}+N_\text{MLBM1,Lag}$ is the total number of Eulerian nodes. We note that for MLBM with $\delta x \ll h$, the type of the interpolation used for a given Lagrangian node is determined by the arrangement of the nodes in the stencil of its target Eulerian point thus we use $N_\text{MLBM1,RBF-FD}$ and $N_\text{MLBM1,Lag}$ in the presented analysis. The factor $3/2$ comes from the arrays of the stencil members indices, usually of \texttt{int} type, which takes half as much memory as a \texttt{double} number. Taking the ratio of the two gives the memory overhead of MLBM1 compared to LBM1
	\begin{equation}\label{eq:ratio_mem}
		r_M=\frac{M_\text{MLBM1}}{M_\text{LBM1}}
	\end{equation}
	Considering that sparser discretization can be used in the meshless algorithm than in LBM1 with the same $\delta x$, MLBM1 can outperform LBM1 regarding memory demands.

	Table~\ref{tab:tau1_memory_demand} shows the number of nodes and the memory needed by LBM1, along with the $r_M$ ratio for the porous medium setups. We note that number of nodes in LBM1 is approximated as $N_\text{LBM1} \approx \varphi \delta x^{-2}$. For example, in the porous medium flow, we used $\delta x \in [2 \cdot 10^{-3}; 1.5 \cdot 10^{-4}]$. In the LBM1 setup with the same streaming distance $\delta x$, it would give the number of nodes ranging from $1.6\cdot 10^5$ for $\delta x = 2 \cdot 10^{-3}$, to $2.84 \cdot 10^7$ for $\delta x=1.5 \cdot 10^{-4}$. At the same time the memory demand of MLBM1 is constant regardless of the value of $\delta x$ and equal to $9.58 \cdot 10^6c$, with $N_\text{MLBM1,RBF-FD} = 26755$ and $N_\text{MLBM1,Lag} = 12190$. This means that for the MLBM1, the ratio $r_M$ ranges from $9.98$ for $\delta x = 2 \cdot 10^{-3}$ to $5.61 \cdot 10^{-2}$ for $\delta x = 1.5 \cdot 10^{-4}$ (or, equivalently, $17.8$ times more memory in LBM1 than in MLBM1). Thus, it is possible to increase the range of Reynolds numbers where $\tau=1$ formulation of LBM exhibits low memory usage by resorting to the interpolation-supplemented, meshless streaming.

	%	JULIA FUNCTIONS USED TO GENERATE THE DATA FOR THE BELOW TABLE:
	%m_mlbm(nscat,nlag) = 2*3*(nscat+nlag) + 3.0/2*(25*nscat + 9*nlag)*8
	%m_lbm(n) = n*6
	%dx =[1.5e-4, 1.8e-4, 2.1e-4, 0.00024, 0.00027, 0.0003, 0.00045, 0.0006, 0.00075, 0.0009, 0.00105, 0.0012, 0.00135, 0.0015, 0.002]
	%n_lbm = @. round(1.0/dx^2 * 0.64) - NUMER OF NODES IN LBM1
	%m_lbm.(n_lbm) - NUMBER OF DOUBLE ADDRESSES IN LBM1
	%m_mlbm(24292+2463,12190) ./ m_lbm.(n_lbm) - r_M RATIO

	\begin{table}[ht!]
		\caption{Memory demands of the LBM1 setups for the porous medium flow for chosen streaming distance lengths $\delta x$ and Reynolds numbers: the number of lattice sites $N_\text{LBM1}$, the corresponding number of \texttt{double} memory addresses $N_\text{LBM1}/c$, and the memory reduction compared to the presented MLBM1 setup $r_M$. The Reynolds numbers and $r_M$ ratios for which MLBM1 uses less memory than LBM1 are given in boldcase.}
				\centering
		\begin{tabular}{ccccc}
			$\delta x$ & $Re$ & $N_\text{LBM1}$ & $M_\text{LBM1}/c$ & $r_M$ \\
			\hline
			$2.00 \cdot 10^{-3}$ & $0.26$ & $1.60 \cdot 10^5$ &  $9.60 \cdot 10^5$ & $9.98$ \\
			$1.35 \cdot 10^{-3}$ & $0.87$ & $3.51 \cdot 10^5$ & $2.10 \cdot 10^6$ & $4.55$ \\
			$1.05 \cdot 10^{-3}$ & $1.84$ & $5.80 \cdot 10^5$ & $3.48 \cdot 10^6$ & $2.75$ \\
			$7.50 \cdot 10^{-4}$ & $5.01$ & $1.14 \cdot 10^6$ & $6.82 \cdot 10^6$ & $1.40$ \\
			$6.00 \cdot 10^{-4}$ & $\mathbf{9.66}$ & $1.78 \cdot 10^6$ & $1.06 \cdot 10^7$ & $\mathbf{0.898}$ \\
			$4.50 \cdot 10^{-4}$ & $\mathbf{21.80}$ & $3.16 \cdot 10^6$ & $1.89 \cdot 10^7$ & $\mathbf{0.505}$ \\
			$2.70 \cdot 10^{-4}$ & $\mathbf{77.57}$ & $8.78 \cdot 10^6$ & $5.26 \cdot 10^7$ & $\mathbf{0.182}$ \\
			$2.10 \cdot 10^{-4}$ & $\mathbf{134.53}$ & $1.45 \cdot 10^7$ & $8.70 \cdot 10^7$ & $\mathbf{0.11}$ \\
			$1.50 \cdot 10^{-4}$ & $\mathbf{262.84}$ & $2.84 \cdot 10^7$ & $1.70 \cdot 10^8 $ & $\mathbf{0.0561}$ \\
			&&& \\
		\end{tabular}
	\end{table}\label{tab:tau1_memory_demand}
%		\centering
%		\begin{tabular}{cccc}
%			$\delta x$ & $N_\text{LBM1}$ & $M_\text{LBM1}/c$ & $r_M$ \\
%			\hline
%			$2.00 \cdot 10^{-3}$ &  $1.60 \cdot 10^5$ &  $9.60 \cdot 10^5$ & $9.98$ \\
%			$1.50 \cdot 10^{-3}$ &  $2.84 \cdot 10^5$ & $1.70 \cdot 10^6$ & $5.61$ \\
%			$1.35 \cdot 10^{-3}$ &  $3.51 \cdot 10^5$ & $2.10 \cdot 10^6$ & $4.55$ \\
%			$1.20 \cdot 10^{-3}$ &  $4.44 \cdot 10^5$ & $2.66 \cdot 10^6$ & $3.59$ \\
%			$1.05 \cdot 10^{-3}$ &  $5.80 \cdot 10^5$ & $3.48 \cdot 10^6$ & $2.75$ \\
%			$9.00 \cdot 10^{-4}$ &  $7.90 \cdot 10^5$ & $4.74 \cdot 10^6$ & $2.02$ \\
%			$7.50 \cdot 10^{-4}$ &  $1.14 \cdot 10^6$ & $6.82 \cdot 10^6$ & $1.40$ \\
%			$6.00 \cdot 10^{-4}$ &  $1.78 \cdot 10^6$ & $1.06 \cdot 10^7$ & $0.898$ \\
%		\end{tabular}
%		\hspace{1cm}
%		\begin{tabular}{cccc}
%			$\delta x$ & $N_\text{LBM1}$ & $M_\text{LBM1}/c$ & $r_M$ \\
%			\hline
%			$4.50 \cdot 10^{-4}$ &  $3.16 \cdot 10^6$ & $1.89 \cdot 10^7$ & $0.505$ \\
%			$3.00 \cdot 10^{-4}$ &  $7.11 \cdot 10^6$ & $4.26 \cdot 10^7$ & $0.224$ \\
%			$2.70 \cdot 10^{-4}$ &  $8.78 \cdot 10^6$ & $5.26 \cdot 10^7$ & $0.182$ \\
%			$2.40 \cdot 10^{-4}$ &  $1.11 \cdot 10^7$ & $6.66 \cdot 10^7$ & $0.144$ \\
%			$2.10 \cdot 10^{-4}$ &  $1.45 \cdot 10^7$ & $8.70 \cdot 10^7$ & $0.11$ \\
%			$1.80 \cdot 10^{-4}$ &  $1.98 \cdot 10^7$ & $1.18 \cdot 10^8$ & $0.0808$ \\
%			$1.50 \cdot 10^{-4}$ &  $2.84 \cdot 10^7$ & $1.70 \cdot 10^8 $ & $0.0561$ \\
%			&&& \\
%		\end{tabular}
%	\end{table}\label{tab:tau1_memory_demand}	
	
	\section{Discussion}
	
	The decoupled space and velocity discretizations in MLBM extend the paradigm of increasing the model's size typical for LBM. In the lattice-based formulation, the information about the characteristic size of the problem is encoded in the number of lattice links/nodes that discretize this size. So is the accuracy of the approximation in space. In MLBM, the information about the system's size is stored in the ratio between the streaming distance and the distance between Eulerian points representing the characteristic length. The approximation accuracy in space depends on the distances between the approximation stencil members. Consequently, in MLBM, the velocity discretization parameters are determined by the desired compressibility errors while the space discretization parameters are determined by the flow domain's geometry and the desired approximation accuracy in space.
	
	The presented results allow us to compare the meshless and lattice-based formulations of LBM and state the following. First, the qualitative results show that the described approach causes the Reynolds number to increase, in line with the findings of previous authors \cite{He97}. Large streaming distances compared to the spacings between the Eulerian points correspond to fast momentum diffusion (low Reynolds number) and small $\delta x/h$ ratios - on the contrary. The quantitative validation of the obtained results suggests that the propagation of information in the model based on the approximation, rather than the exact streaming, does not impair the physical mechanism of mass and momentum transfer in the lattice Boltzmann equation on scales relevant to the studied phenomena.
	
	The accuracy and convergence of MLBM and other off-grid LBM methods were investigated previously~\cite{Kramer2020,Strzelczyk2024}, especially in application to inertial flows in works by Lin and others \cite{Lin2019} or Musavi and Ashrafizaadeh \cite{Musavi2019,Musavi2016}. From the point of view of the solution's error, the decrease of the streaming distance length $\delta x$ acts in favor of the compressibility errors. In our simulations, the decreasing $\delta x$ allows us to achieve higher physical values of the inlet velocity and the body force density, which are constant in lattice units. At the same time, the decreasing ratio of $\delta x/h$ introduces larger approximation errors in OLBM~\cite{Kramer2020}. When semi-Lagrangian streaming is used, this has directly to do with the error term $\mathcal{O}\left(h^{p+1}/\delta x\right)$ characteristic for semi-Lagrangian methods, as suggested in our previous work \cite{Strzelczyk2024}. So, ultimately, increasing the Reynolds number by decreasing the $\delta x/h$ ratio will lead to the instability of the solution due to the approximation errors. The error of the solutions obtained with MLBM at low $\delta x/h$ may also come from the under-resolution of physical phenomena, such as vortices typical for inertial flows. However, this under-resolution might be exploited to the method's advantage, as suggested by Chen \cite{Chen1998}. The most straightforward way to recover the desired approximation accuracy is to increase the Eulerian points density or use higher-order approximations. Unfortunately, both of those approaches lead to the rise of computational demands of the model.
	
	To avoid computational overhead, one can exploit the two following features of the meshless approximation methods. First, the feasibility of the local refinement, which allows to vary the internodal distance in space. The presented results used arbitrary refinement, based on the known or assumed solution of the transport equations, and were shown to grant a several-fold reduction in execution time compared to the uniform discretization. Second, the savings may be achieved by switching to regular discretizations with the approximation schemes computationally less intensive than the ones using scattered node sets. The research on this approach is given attention in the field of meshless analysis~\cite{Rot2024,Javed2013,Ding2004} and our results show it is possible to reduce the execution time by about $40\%$ this way, compared to a fully scattered node set. We hypothesize that the two approaches together give especially high complexity reduction in flows where clearly distinguished length scales occur, e.g. a small obstacle inside a large bulk of fluid. Complimentary to refining the discretization and using hybrid (regular-irregular) points arrangements \textit{a priori}, the choice of a proper error indicator, such as quantities based on vorticity suggested by Fakhari and Lee \cite{Fakhari2014} or on the velocity field divergence \cite{Succi2018}, can allow for dynamic $h$- or $p$-refinement during the simulation run. In turn, the error of the solution becomes more uniform in space granting a better accuracy at a little increase of the computational expense.
	
	Finally, the fixed-relaxation time models~\cite{Matyka2021,Zhou2020} reduce the memory overhead by explicitly storing just a few macroscopic fields rather than a large number of velocity distributions. We show that by introducing hybrid space discretization, it is possible to achieve a several-fold memory advantage in comparison to the lattice-based LBM1, which is already estimated to use about $76\%$ less memory than its arbitrary relaxation time counterpart \cite{Matyka2021}. This comes at the cost of the increased computational burden due to the approximation step, which we estimate to be higher even than in the case of the meshless arbitrary-relaxation time LBM, on the contrary to the lattice-based schemes (see Appendix~\ref{app:computational_expense}).
	
	\section{Conclusions}
	
	We investigate the procedure to increase the Reynolds number of a flow from creeping to transitional regime in meshless LBM simulations by decreasing the streaming distance, such that no additional nodes are inserted into the domain. We apply this method of the Reynolds number increase to a general meshless LBM scheme and to the meshless implementation of a memory-efficient LBM model with the fixed relaxation time $\tau=1$ (LBM1). We show that the procedure works by numerically investigating two test cases -- von K\'arm\'an vortex street behind a cylindrical obstacle of infinite length and a flow through a porous sample -- in a range of Reynolds numbers. The qualitative observation of the velocity fields indicates the onset of inertial effects in the flows. Also, the obtained values of hydrodynamic parameters agree well with the numerical and experimental references for the considered Reynolds numbers. We highlight approximately $12$-fold computational speedup of MLBM compared to the standard LBM when a proper choice of the nodes arrangement and approximation scheme is made. From the performed simulations, we estimate the possible memory savings of the meshless LBM1 compared to the standard (lattice-based) LBM1 to be as high as $17$-fold and pinpoint the source of the computational overhead of MLBM1. Finally we note that complimentary to optimizing approximation setup in the MLBM models, stability improvements can be obtained by using more stable collision operators, such as central moment LBM~\cite{Ning2016}.

	\section{Acknowledgments}
	Funded by National Science Centre, Poland under the OPUS call in the Weave programme 2021/43/I/ST3/00228. This research was funded in whole or in part by National Science Centre (2021/43/I/ST3/00228). For the purpose of Open Access, the author has applied a CC-BY public copyright licence to any Author Accepted Manuscript (AAM) version arising from this submission.
	
	\appendix
	
	\section{Comparison of computational demands of lattice-based and meshless LBM}\label{app:computational_expense}
	
	The computational complexity of MLBM and MLBM1 carries the overhead of interpolation during the streaming step. Table~\ref{tab:lbms_complexity} shows the number of operations performed in each point in the discretization per timestep for the four discussed LBM models. $O_{col}$ corresponds to the collision step, Eq.~\eqref{eq:LBM_collision}, $O_{eq}$ corresponds to the calculation of the equilibrium VDF, Eq.~\eqref{eq:feq}, $O_{int}$ is the number of operations performed during the interpolation of one scalar, and $O_m$ is the number of operations performed to add the contribution of one population to all the macroscopic moments. Comparing MLBM1 to LBM1, the computational overhead of the former comes from the interpolation of $N_f$ fields to all but one departure node. On the other hand, when MLBM1 is compared with MLBM, the $N_f$ times more interpolations in the former is traded for the calculation of $q$ post-collision VDFs in the latter. Considering that $O_{int} \sim \mathcal{O}(N_L)$, the product $N_fO_{int} \gg O_{col}$ in models like SRT or TRT. Thus, in MLBM1, the simpler collision term, than in MLBM, cannot make up for the additional approximation burden. Finally, one observes that in the lattice-based formulations (LBM and LBM1), fixing the relaxation time slightly reduces the computational complexity of the scheme (by the term $qO_{col}$), while in the meshless ones increases it significantly (by the term $(q-1)(N_f-1)O_{int} - qO_{col}$).
	\begin{table}[!ht]
		\caption{Number of operations performed for each point in the discretization per timestep for the LBM models considered in this work.}
		\centering
		\begin{tabular}{rcccc}
			& \textbf{LBM} & \textbf{LBM1} & \textbf{MLBM} & \textbf{MLBM1} \\
			\textbf{approximation} & $0$ & $0$ & $(q-1)O_{int}$ & $(q-1)N_fO_{int}$ \\
			\textbf{equilibrium calculation} & $qO_{eq}$ & $qO_{eq}$ & $qO_{eq}$ & $qO_{eq}$ \\
			\textbf{collision} & $qO_{col}$ & $0$ & $qO_{col}$ & $0$ \\
			\textbf{moments update} & $qO_m$ & $qO_m$ & $qO_m$ & $qO_m$ \\
		\end{tabular}
	\end{table}\label{tab:lbms_complexity}
	
	\bibliography{main.bbl}

\end{document}